\documentclass[usenatbib]{mnras}
\defcitealias{Sironi_20}{SB20}
\defcitealias{Sridhar+21c}{SSB21}

\usepackage[T1]{fontenc}
\usepackage{ae,aecompl}
\usepackage{cancel}

\usepackage{graphicx}	% Including figure files
\usepackage{amsmath}	% Advanced maths commands
\usepackage{amssymb}	% Extra maths symbols
\usepackage{threeparttable}
\usepackage{booktabs,caption}
\usepackage{changepage}
\usepackage{adjustbox}
\usepackage{graphics}
\usepackage{subcaption}
\usepackage{mathrsfs}
\usepackage{lipsum,verbatim}

\newcommand{\ux}{u_{\rm x}}

\newcommand{\gcr}{\gamma_{\rm cr}}
\newcommand{\gammacr}{\gamma_{\rm cr}}
\newcommand{\gammae}{\gamma_{\rm e}}
\newcommand{\comp}{c/\omega_{\rm p}}
\newcommand{\compe}{c/\omega_{\rm pe}}
\newcommand{\compi}{c/\omega_{\rm pi}}
\newcommand{\sigmai}{\sigma_{\rm i}}
\newcommand{\sigmae}{\sigma_{\rm e}}
\newcommand{\mi}{m_{\rm i}}
\newcommand{\me}{m_{\rm e}}
\newcommand{\gcool}{\tau_{\rm cool}}
\newcommand{\Lx}{L_{\rm x}}

\newcommand{\eqb}{\begin{eqnarray}}
\newcommand{\eqe}{\end{eqnarray}}
\newcommand{\be}{\begin{eqnarray}}
\newcommand{\ee}{\end{eqnarray}}
\newcommand{\bi}{\begin{itemize}}
\newcommand{\ei}{\end{itemize}}
\newcommand{\fHE}{f_{\rm HE}}
\title[Cold-chain Comptonization in Black Hole Coronae II]{Comptonization by Reconnection Plasmoids in Black Hole Coronae II: Electron-Ion Plasma}

\author[N. Sridhar, L. Sironi \& A. Beloborodov]{
Navin Sridhar,$^1$\thanks{E-mail: navin.sridhar@columbia.edu}
Lorenzo Sironi,$^1$\thanks{E-mail: lsironi@astro.columbia.edu} and
Andrei M. Beloborodov$^{2,3}$\thanks{E-mail: amb2046@columbia.edu }
\\
$^1$Department of Astronomy and Columbia Astrophysics Laboratory, Columbia University, 550 W 120th St, New York, NY 10027, USA\\
$^2$Department of Physics and Columbia Astrophysics Laboratory, Columbia University, 550 W 120th St, New York, NY 10027, USA\\
$^3$Max Planck Institute for Astrophysics, Karl-Schwarzschild-Str. 1, D-85741, Garching, Germany
}

\date{Accepted XXX. Received YYY; in original form ZZZ}

\pubyear{2020}

\begin{document}
\label{firstpage}
\pagerange{\pageref{firstpage}--\pageref{lastpage}}
\maketitle

\begin{abstract}
We perform two-dimensional particle-in-cell simulations of magnetic reconnection in electron-ion plasmas subject to strong Compton cooling and calculate the X-ray spectra produced by this process. The simulations are performed for trans-relativistic reconnection with magnetization $1\leq \sigma\leq 3$ (defined as the ratio of magnetic tension to plasma rest-mass energy density), which is expected in the coronae of accretion disks around black holes. We find that magnetic dissipation proceeds with inefficient energy exchange between the heated ions and the Compton-cooled electrons. As a result, most electrons are kept at a low temperature in Compton equilibrium with radiation, and so thermal Comptonization cannot reach photon energies $\sim 100\,$keV observed from accreting black holes. Nevertheless, magnetic reconnection efficiently generates $\sim 100\,$keV photons because of mildly relativistic bulk motions of the plasmoid chain formed in the reconnection layer. Comptonization by the plasmoid motions dominates the radiative output and controls the peak of the radiation spectrum $E_{\rm pk}$. We find $E_{\rm pk}\sim 40$\,keV for $\sigma=1$ and $E_{\rm pk}\sim100$\,keV for $\sigma=3$. In addition to the X-ray peak around 100\,keV, the simulations show a non-thermal MeV tail emitted by a non-thermal electron population generated near X-points of the reconnection layer. The results are consistent with the typical hard state of accreting black holes. In particular, we find that the spectrum of Cygnus~X-1 is well explained by electron-ion reconnection with $\sigma\sim 3$.
\end{abstract}


%%%%%%%%%%%%%%%%%%%%%%%%%%%%%%%%%%%%%%%%%%%%%%%%%%

%%%%%%%%%%%%%%%%% BODY OF PAPER %%%%%%%%%%%%%%%%%%

\section{Introduction}

Hard, non-thermal X-rays observed from black hole X-ray binaries (BHXBs) are traditionally attributed to thermal Comptonization of seed soft photons by a hot plasma (`corona') surrounding the black hole (BH), with typical inferred electron temperatures $\sim$100\,keV \citep{Zdziarski&Gierlinski_04}. X-ray spectral and temporal studies have shed light on how the geometry and location of the corona change depending on the phase of the outburst \citep{Kara_19, Sridhar+20, Wang+21, Cao+21}. Yet, the mechanism that energizes the corona against the strong inverse Compton (IC) 
losses is still a mystery. Recently, \cite{belo_17} suggested that magnetic reconnection can power the Comptonization process through the bulk motions of plasmoids in the reconnection layer.\footnote{See Figure~1 in \cite{Sridhar+21c} for a schematic illustration of the proposed `cold-chain Comptonization' process.} The plasmoids sustain relativistic speeds if the magnetic energy density $B^2/8\pi$ exceeds the plasma rest mass energy density $\rho c^2$, which corresponds to a magnetization parameter $\sigma\equiv B^2/4\pi\rho c^2 > 1$ The Comptonization model of \cite{belo_17} was followed by full kinetic simulations of reconnection in $e^\pm$ plasma with strong IC cooling (\citealt{werner_19}; \citealt{Sironi_20}, hereafter \citetalias{Sironi_20}; \citealt{Sridhar+21c}, hereafter \citetalias{Sridhar+21c}). The simulations (\citetalias{Sironi_20}, \citetalias{Sridhar+21c}) confirmed that the pair plasma trapped in plasmoids is efficiently cooled to a non-relativistic temperature while their bulk motions remain trans-relativistic, resembling a Maxwellian distribution with effective temperature $k_{\rm B}T_{\rm bulk}\sim100$\,keV. Furthermore, Monte Carlo simulations of X-ray spectra produced by the `cold-chain Comptonization' approximately agreed with the observed hard state of BHXBs in particular Cygnus X-1 \citep{Sridhar+21c}. Our previous numerical simulations \citetalias{Sironi_20}, \citetalias{Sridhar+21c}) focused on reconnection in strongly magnetized ($\sigma\gg1$) $e^\pm$ pair plasma. However, accretion disk coronae can also have ion-dominated regions with $\sigma\sim 1$ adjacent to the accretion flow,
\citep[e.g.,][]{Galeev+79}. Formation of disk coronae is a difficult problem studied with GRMHD simulations \citep[e.g.,][]{jiang_19, chatterjee_19,ripperda_20,ripperda_21,Nathanail+21}. It is not established
which region dominates the observed hard X-ray emission. It is possible that magnetic reconnection in electron-ion plasma with moderate $\sigma\sim 1$ is the dominant source of hard X-rays.

The present paper investigates this possibility using radiative kinetic simulations. Similar simulations of trans-relativistic reconnection in electron-ion plasmas 
were performed before only in the absence of radiative cooling \citep{Melzani+14a, Melzani+14b, rowan_17, werner_18, ball_18,rowan_19, Ball+19, Kilian+20}. We, for the first time, extend these studies to the radiative regime. In particular, we investigate the effect of IC losses on the particle energy distribution and evaluate the IC emission from the reconnection layer.

Only electrons experience significant IC cooling. As a result, the reconnection layer can enter a two-temperature state with ions much hotter than the electrons. This depends on the efficiency of the electron-ion energy exchange. Exchange through Coulomb collisions is typically slow in the corona with $\sigma>1$, however there may be collisionless mechanisms---involving e.g. ion velocity-space instabilities \citep{Hasegawa+69,  Gary_92, Sharma+06, Kunz+14, Riquelme+15, Sironi&Narayan_15, zhdankin_20b}---that could transfer the ion thermal energy to the colder electrons.  This suggests a possibility for sustaining a high electron temperature $T_{\rm e}$ inside reconnection plasmoids, perhaps competitive with $T_{\rm bulk}$ of the plasmoid bulk motions. Then, thermal Comptonization could become important for hard X-ray production.  Kinetic plasma simulations are the best way to find out, from first principles, the importance of $T_{\rm e}$ vs. $T_{\rm bulk}$.
\footnote{In paper I (\citetalias{Sridhar+21c}), we presented an experiment where only electrons were selectively cooled, with positrons remaining hot---mimicking a radiatively-cooled electron-ion plasma. Our results showed that no energy is efficiently transferred from the hot positrons to the colder electrons, on timescales shorter the plasma advection time out of the reconnection layer.} The simulations will also show how the electron distribution in the reconnection layer depends on $\sigma$ and what effect it has on the Comptonization spectrum.

The paper is organized as follows. In \S\ref{sec:setup}, we describe the numerical setup of our simulations. In \S\ref{sec:IC_cooling}, we present the implementation and parameterization of IC cooling, and the most important timescales for our problem, in the context of BH coronae. In \S\ref{sec:results}, we present our PIC results, emphasizing the dependence on magnetization and strength of IC cooling. The photon spectra derived from our PIC simulations using Monte Carlo radiative transfer calculations are presented in \S\ref{sec:radiative_transfer}. We summarize our findings in \S\ref{sec:summary}.

\section{PIC simulation setup} \label{sec:setup}

We perform simulations of magnetic reconnection with the 3D electromagnetic particle-in-cell code {\sc tristan-mp} \citep{buneman_93, spitkovsky_05}; the code employs a Vay pusher \citep{Vay_08} for updating the particles' momenta. Overall, the simulation setup is similar to what we have employed in paper I of this series (\citetalias{Sridhar+21c}) and in \citetalias{Sironi_20}. We nonetheless describe it here given the different composition of the plasma.  We refer readers to Appendix~\ref{appendix:table} for the complete set of our numerical and physical input parameters.

Our simulations are performed on a 2D spatial domain---with Cartesian grid in the $x{-}y$ plane---yet evolving all three components ($x,y,z$) of velocities and electromagnetic fields. We initialize a plasma composed of electrons (subscript `e') and protons/ions (subscript `i'). The effect of the proton-to-electron mass ratio is examined in Appendix~\ref{appendix:mass_ratio} for $\mi/\me\simeq7-1836$; our fiducial mass ratio is $\mi/\me=29$, which in the uncooled regime gives results in reasonable agreement with realistic mass ratio (see Appendix~\ref{appendix:mass_ratio}). Electrons and ions have equal number densities, $n_{\rm 0e}=n_{\rm 0i}$, and can be described by a Maxwell-J\"{u}ttner distribution with dimensionless temperatures $\theta_{\rm e} = k_{\rm B}T_{\rm e}/m_{\rm e} c^2 \sim 10^{-4}$ and $\theta_{\rm i} = k_{\rm B}T_{\rm i}/m_{\rm i} c^2 = 10^{-4}(\me/\mi)$, respectively, where we initialize electrons and ions with the same temperature $T_{\rm e}=T_{\rm i}$. The Debye length corresponding to these conditions is $\lambda_{\rm De}=\sqrt{k_{\rm B}T_{\rm e}/4\pi n_0 e^2}=0.05$ cells. We demonstrate in Appendix~\ref{appendix:resolution_convergence} (on simulation convergence with spatial resolution) that this choice of upstream temperature does not influence our overall conclusions, by performing additional simulations with $\theta_{\rm e}=0.02$, or equivalently, $\lambda_{\rm De}=1$ cell.

For our case of cold plasma, we define the plasma magnetization of a given species $\sigma_{\rm s}$\footnote{Throughout this paper, the subscript `s' can denote either species---electrons or ions; the subscript `$e^\pm$' is used to denote electrons and positrons in a pair plasma. E.g., when referring to results from paper I (\citetalias{Sridhar+21c}), the magnetization $\sigma_{e^\pm}$ is normalized to the electron-positron mass density.} as:
\begin{equation} \label{eqn:magnetization}
\sigma_{\rm s}=\dfrac{B^2_{\rm 0}}{4\pi n_{\rm 0s}m_{\rm s}c^2} = \bigg(\dfrac{\omega_{\rm cs}}{\omega_{\rm ps}}\bigg)^2,
\end{equation}
where $B_{\rm 0}$ is the asymptotic upstream magnetic field strength, $\omega_{\rm cs}=eB_{\rm 0}/m_{\rm s}c$ is the particle gyro-frequency, the plasma frequency $\omega_{\rm ps}$ is defined as
\begin{equation} \label{eqn:plasma_freq}
\omega_{\rm ps}=\sqrt{\frac{4\pi n_{\rm 0s} e^2}{m_{\rm s}}},
\end{equation}
with the corresponding plasma skin depth given by $c/\omega_{\rm ps}$. The plasma we initialize is dominated in mass by the ions, therefore the overall plasma's magnetization parameter $\sigma\equiv B^2/4\pi\rho c^2$ is nearly equal to $\sigmai$. 
We are interested in the regime of trans-relativistic reconnection, so we consider $\sigmai=\sigmae\me/\mi=\{1,2,3\}$. The corresponding Alfv\'{e}n speeds are $v_{\rm A}/c = \sqrt{\sigmai/(\sigmai+1)} = \{0.71,0.81,0.87\}$. The plasma-$\beta$ parameter of the initialized plasma is given by $\beta_{\rm 0}=\beta_{\rm i}+\beta_{\rm e}=8\pi k(n_{\rm i}T_{\rm i} + n_{\rm e}T_{\rm e})/B_{\rm 0}^2 \sim \theta_{\rm s}/\sigma_{\rm s}$. Our choices of the upstream plasma temperature and magnetization always renders the plasma $\beta_{\rm 0}\ll1$.

The fields undergoing reconnection are initialized along the $\pm\hat{x}$ direction ($B_x$), while the reconnected fields get oriented along the $\pm\hat{y}$ direction. This sets the plasma inflow and outflow directions to be along $\pm\hat{y}$ and $\pm\hat{x}$, respectively. We initialize the system in the so-called Harris equilibrium \citep{1962NCim...23..115H}, where the initial magnetic field profile is given by $B_x=-B_{\rm 0}\tanh(2\pi y/\Delta_{\rm y})$, i.e., at $y=0$ the field reverses its polarity over a thickness of $\Delta_{\rm y}$. Consistent with our previous work (\citetalias{Sridhar+21c}), we also add a uniform guide field $B_{\rm g}=0.1\,B_{\rm 0}$ along the $z$-axis, which does not get dissipated, and so does not add any free energy to the system.

The magnetic pressure outside the layer is balanced by particle pressure in the layer, by adding an additional population of hot particles with an overdensity of 3. We initiate reconnection by artificially cooling down the hot particles near the center [$(x,y)=(0,0)$] at the initial time. We restrict the horizontal extent of the region where we trigger reconnection to a length $\Delta_{\rm x}=400\,\compe$. The reconnected field lines form an `X'-point at the center of the domain, which eventually separates into two reconnection fronts that carry away the plasma along $\pm\hat{x}$ at $\sim$Alfv\'{e}nic speeds. The initial thickness of the current sheet is chosen to be large enough ($\Delta_{\rm y} \simeq 150\,\compe$) such that the tearing mode instability does not spontaneously grow before the two reconnection fronts have reached the boundaries of the domain. 

We employ outflow boundary conditions along the $x$ boundaries of the computational domain \citep{daughton_06, belyaev_15, cerutti_15}. This ensures that the hot plasma initialized in the current sheet gets evacuated after $\sim \Lx/v_{\rm A}$, where $\Lx$ is the half-length of the box along $x$. Along  the $y$ direction, we employ two injectors that introduce fresh plasma and magnetic flux into the computational domain. These injectors move away from $y=0$ at the speed of light, and the box dynamically expands when they reach its boundaries \citep[see][for details]{sironi_16}. With this setup, the reconnection dynamics evolves in a ``quasi-steady state" at time $t>\Lx/v_{\rm A}$ i.e., independent of the inital conditions (viz., temperature or overdensity of hot plasma in the sheet, $\Delta_{\rm x}$ and $\Delta_{\rm y}$). We demonstrate this in Appendix~\ref{appendix:recrate} by showing how the reconnection rate ($\eta_{\rm rec}$; defined as the plasma's inflow velocity in units of the Alfv\'{e}n speed) evolves with time. Akin to \citetalias{Sridhar+21c}, we evolve the simulation up  to $t_{\rm sim}\sim 5 \Lx/v_{\rm A}$ for all the considered magnetizations; this provides a sufficiently long time to assess the steady-state properties of the reconnection system.

We choose our computational domains with different sizes: $420\le \Lx/(\compe)\le 6720$, or equivalently, $78\le \Lx/(\compi)\le 1259$ (for $\mi/\me=29$). Large boxes are essential to retain the correct hierarchy of time and energy scales expected for BH coronae (see \S\ref{sec:IC_cooling}). We resolve the electron skin depth $\compe$ with 5 grid lengths ($\delta$) (i.e., resolution ${\cal R}=(\compe)/\delta=5$). In Appendix~\ref{appendix:resolution_convergence}, we show for a representative case that our results are qualitatively the same with lower and higher resolutions (${\cal R}=2.5, 10$). The numerical speed of light (the Courant-Friedrichs-Lewy number) is set to 0.45 cells/timestep. We initialize the plasma with a density of $n_0=4$ particles per cell (2 per species), and improve particle noise in the electric current density with 32 passes of a ``1-2-1'' low-pass digital filter \citep{birdsall_91} at each step of the PIC loop. We present a convergence test in Appendix~\ref{appendix:ppc_ntimes_convergence}, where we compare our results with simulations having a higher initial particle number density, $n_0=64$, and a lower number of filtering passes (8).

\section{Inverse Compton cooling}\label{sec:IC_cooling}

For each particle in our simulation with velocity $\mathbf{v_{\rm s}}$ ($=\boldsymbol{\beta}_{\rm s}c$) and energy $\gamma_{\rm s} m_{\rm s}c^2$, the IC cooling experienced by it---due to an isotropic bath of soft photons---is implemented in our code with a ``drag force'' \citep{2010NJPh...12l3005T} given by
\begin{equation} \label{eqn:f_ic}
{\mathbfit F}_{\rm IC,s} = -\frac{4}{3}\sigma_{\rm Ts}\gamma_{\rm s}^2U_{\rm rad} \boldsymbol{\beta}_{\rm s},
\end{equation} 
where $\sigma_{\rm Ts}=8\pi e^4/(3m_{\rm s}^2 c^4)$ is the Thomson cross section, $\gamma_{\rm s}=(1-\beta_{\rm s}^2)^{-1/2}$ is the particle Lorentz factor, and $U_{\rm rad}$ is the radiation energy density. Since $\sigma_{\rm Ti}/\sigma_{\rm Te}=\me^2/\mi^2$ and $\gamma_{\rm i}/\gamma_{\rm e}\sim\sigmai/\sigmae=\me/\mi$, the ratio of the drag force experienced by ions and electrons is $F_{\rm IC,i}/F_{\rm IC,e}\simeq\me^4/\mi^4\ll1$, even for our reduced mass ratio $\mi/\me\sim29$. Thus, we restrict the following discussion to electrons.  

We parametrize the radiation energy density $U_{\rm rad}$ in the code by defining a critical Lorentz factor $\gcr\equiv \sqrt{3e\eta_{\rm rec}B_0/(4\sigma_{\rm Te}U_{\rm rad})}$, at which the Compton drag force experienced by the electrons is balanced by the force due to the electric field associated with magnetic reconnection, assuming a reconnection rate $\eta_{\rm rec}\sim 0.1$ \citep{Cassak+17}.\footnote{The electric field associated with reconnection is given by $E_{\rm rec}=\eta_{\rm rec}(v_{\rm A}/c)B_0$. When defining $\gcr$, we take $E_{\rm rec}\sim\eta_{\rm rec}B_0$, which is a fair assumption for our case of trans-relativistic reconnection with $\sigma\sim1$.}
We investigate the reconnection dynamics for the case of no IC cooling ($\gcr=\infty$), and for increasing strengths of IC cooling: $20\gtrsim\gcr\gtrsim7$. Given our simulations' temporal resolution ($\Delta t = 0.09\, \omega_{\rm pe}^{-1}$), the choice of $\gcr > \sqrt{\sigma_{\rm e}}$ ensures that the IC cooling time for any electron---with Lorentz factors even as high as $\gammae\sim\gcr$---is well resolved (\citetalias{Sironi_20}, \citetalias{Sridhar+21c}). 

We refer readers to part I in this series of papers (\citetalias{Sridhar+21c}) and \citetalias{Sironi_20} for a detailed quantification of the plasma conditions in black hole coronae. For an $e^\pm$ corona, we showed that the plasma magnetization is expected to be $\sigma_{e^\pm}\sim400$. On the other hand, the coronae adjacent to the accretion disk can have moderately-magnetized regions ($\sigma\sim1$) whose mass density is dominated by ions---as also seen in GRMHD simulations \citep{chatterjee_19, jiang_19}. It follows that  $\sigmae=\sigmai(\mi/\me)\gg1$. On the other hand, the parametrization for the seed photon density $U_{\rm rad}$, and so $\gcr$, does not depend on the plasma composition, and is estimated to be $\gcr\sim10^4(M_{\rm BH}/M_{\odot})^{1/4}$, where $M_{\rm BH}$ is the mass of the black hole \citetalias{Sridhar+21c}. 

We showed in our earlier works (\citetalias{Sironi_20}, \citetalias{Sridhar+21c}) that electrons accelerated at X-points typically attain Lorentz factors $\gamma_{\rm X}\sim\sigma_{\rm e}/8$,\footnote{This estimate assumes that half of the available magnetic energy is dissipated into particle energy \citep{sironi_15}, and that  electrons and ions acquire the same amount of dissipated field energy.}
and that their IC cooling timescale is much longer than the X-point acceleration timescale. This sets the requirement of $\gcr \gg \sigma_{\rm e}/8$, which ensures that electrons accelerated at the X-points do not suffer significant IC cooling. On the other hand, the IC cooling timescale is  much shorter than the timescale for plasma advection across the reconnection region, whose size, in BH coronae, is comparable to the gravitational radius of the black hole \citep{belo_17}. The ratio of these two timescales in our simulations can be cast as follows (\citetalias{Sridhar+21c}):
\begin{equation} \label{eqn:varrho}
\gcool \equiv \frac{\gcr^2}{\eta_{\rm rec}\sqrt{\sigmae}}\frac{c/\omega_{\rm pe}}{\Lx} < 1~.
\end{equation}
Eq.~\ref{eqn:varrho} is satisfied for large enough domains; we nonetheless explore both the strongly cooled ($\gcool<1$) and weakly cooled ($\gcool>1$) regimes. Values of $\gcool\ll1$ ensure that electrons have enough time to cool down to non-relativistic temperatures before being advected out of the box. On the other hand, in the regime $\gcool\gg1$, electrons exit the domain with a Lorentz factor $\gammae\sim\gcool$, and approach the asymptotic value $\gammae\rightarrow\sigmae/8$ for $\gcool\gtrsim\sigmae$. Throughout our study, we employ $\gcool$ to quantify the strength of radiative cooling experienced by the particles.

\section{PIC simulation results} \label{sec:results}

In this section, we present the results from our PIC simulations of radiative reconnection, and discuss how the presence of ions in the plasma modifies our findings as compared to radiative reconnection in an $e^\pm$ plasma (\citetalias{Sironi_20}, \citetalias{Sridhar+21c}).

In \S\ref{subsec:general}, we present the general structure of the reconnection region for different values of plasma magnetization, exposed to the same level of radiative cooling (same $\gcool$). In \S\ref{subsec:motions}, we demonstrate how the particles' internal and bulk motions are influenced by the plasma magnetization and the level of IC cooling. The resulting energy spectra of  electrons---separated into their internal and bulk motions---are then presented in \S\ref{subsec:spectra}.

\subsection{Characteristics of the reconnection region} \label{subsec:general}

\begin{figure}
\includegraphics[width=8cm]{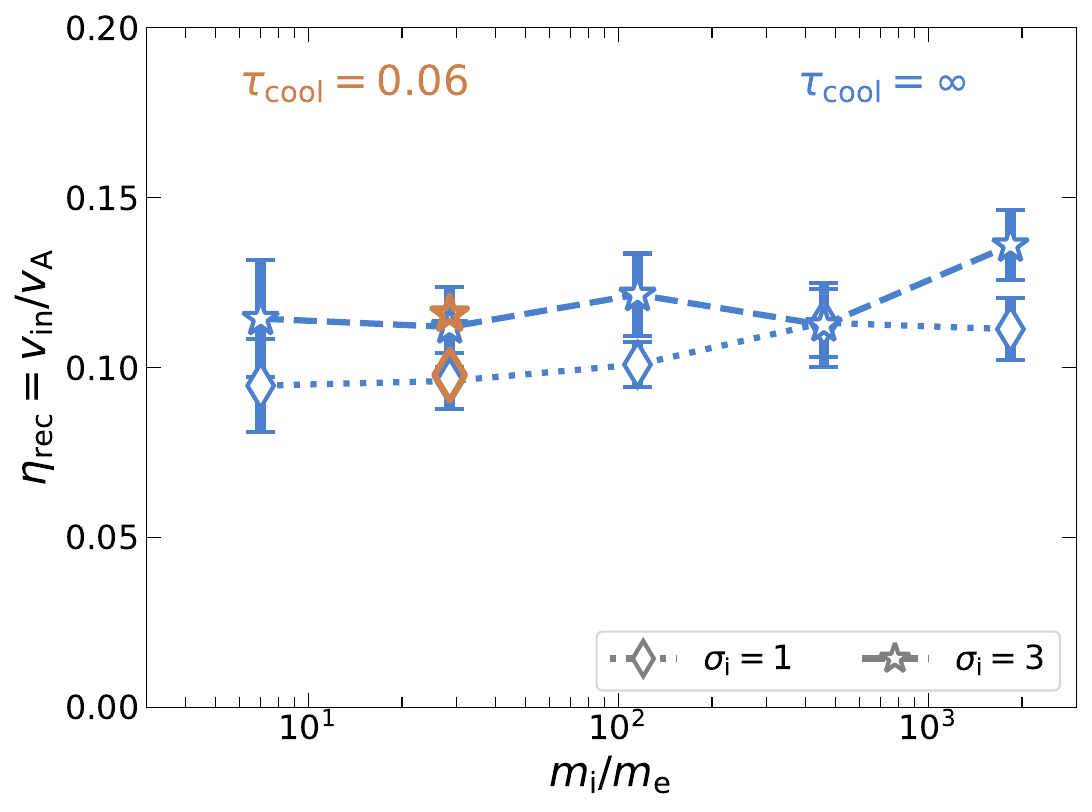}
\caption{Reconnection rate $\eta_{\rm rec}$, defined as the speed of the plasma inflow towards the reconnection region ($v_{\rm in}$) normalized by the Alfv\'{e}n speed ($v_{\rm A}$). The dotted and dashed lines connecting the diamond and star markers, respectively, present the dependence of $\eta_{\rm rec}$ on $\mi/\me$ for $\sigmai=1$ and 3, respectively. The blue markers denote uncooled cases ($\gcool=\infty$), whereas the brown markers show the results of the strongest cooling case ($\gcool=0.06$), for our fiducial $\mi/\me=29$. The inflow rate is spatially-averaged over a rectangular slab located above the reconnection mid-plane (at $0.1<x/\Lx<0.9$ and $0.05<y/\Lx<0.15$, where $\Lx/(\compe)=1680$), and is then averaged during the time interval $2 \lesssim T/(\Lx/v_{\rm A}) \lesssim 5$. The error bars denote the standard deviation.}
\label{fig:recrate}
\end{figure}

\begin{figure*}
    \centering 
    \begin{subfigure}[t]{0.5\textwidth}
        \centering
        \includegraphics[trim={2cm 0 2cm 0},clip,height=3.4in]{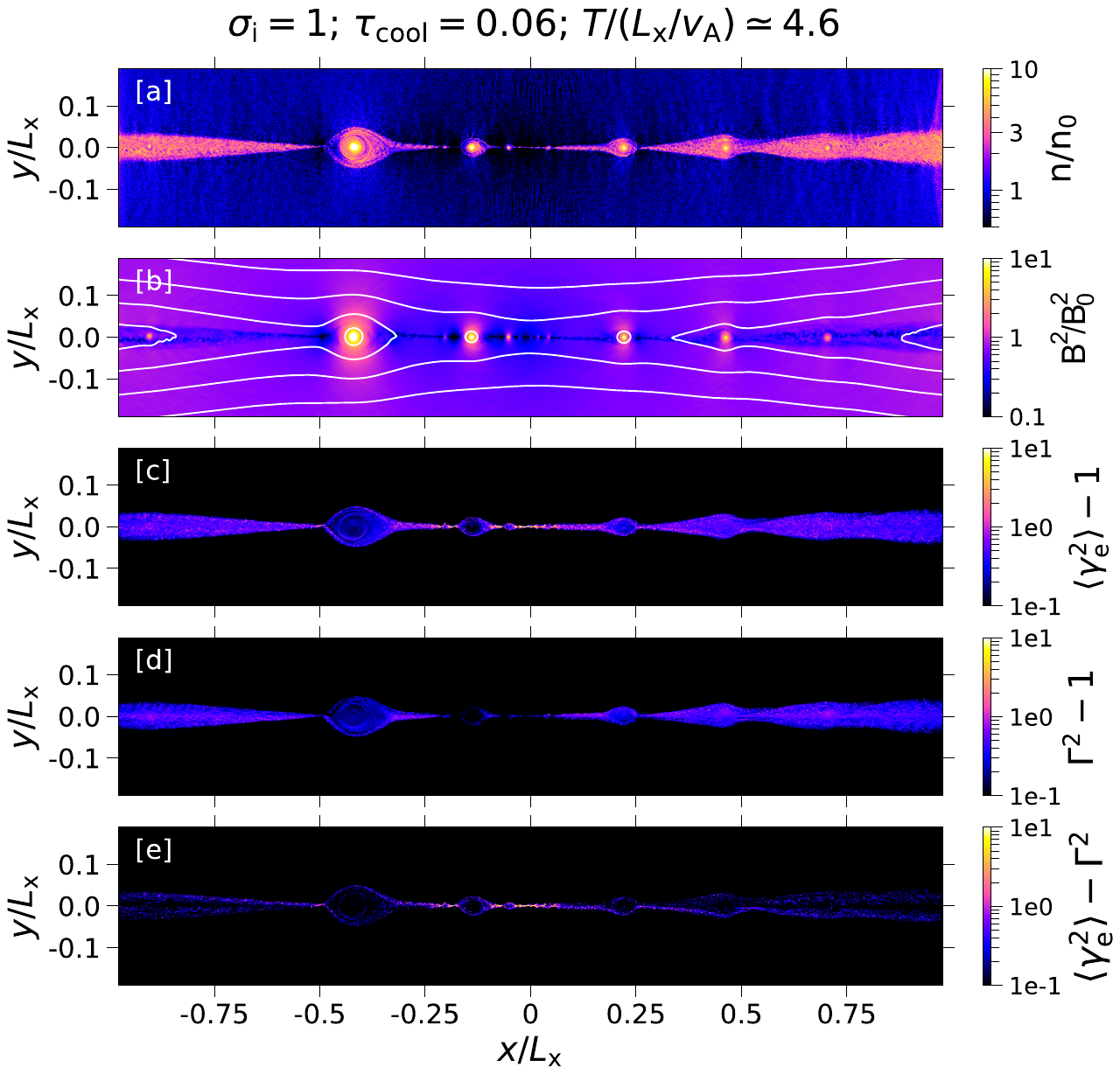}   
        \caption*{(i) $\sigmai=1$}
    \end{subfigure}%
    ~ 
    \begin{subfigure}[t]{0.5\textwidth}
        \centering
        \includegraphics[trim={2cm 0 2cm 0},clip,height=3.4in]{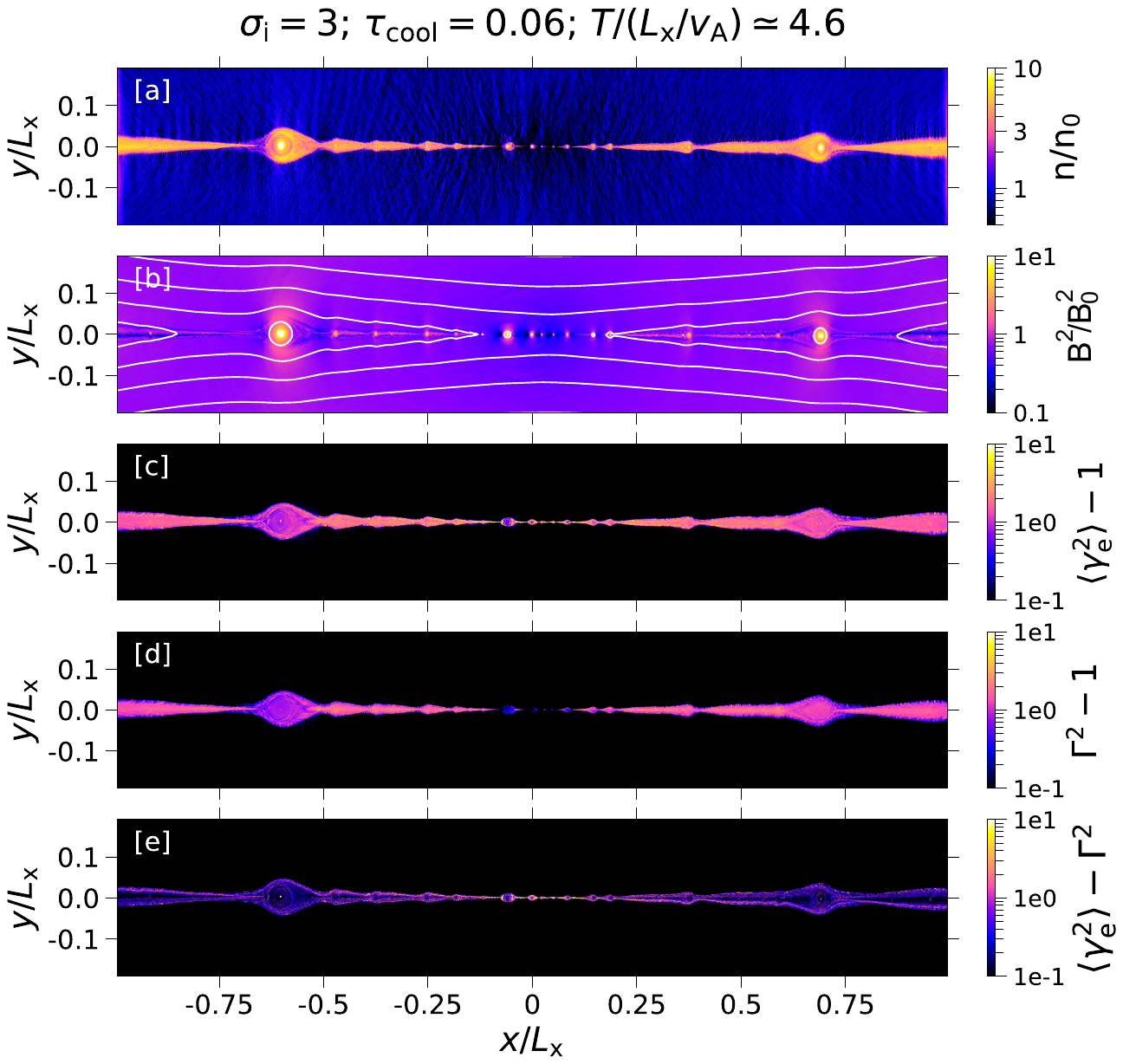}   
        \caption*{(ii) $\sigmai=3$}
    \end{subfigure}
    \caption{Snapshot of the reconnection region at time $T\simeq 4.6\,\Lx/v_{\rm A}$ for two magnetization models: $\sigmai=1$ (left) and $\sigmai=3$ (right), with otherwise the same level of cooling ($\gcool=0.06$) and mass ratio ($\mi/\me=29$; see Appendix~\ref{appendix:table} for other runs). We only show the region $|y|\lesssim 0.2\,\Lx$ where reconnection occurs, where $\Lx/(\compe)=1680$ and 6720 for the $\sigmai=1$ and 3 models, respectively. [a] Particle number density $n$ in units of the initialized (upstream) number density $n_{\rm 0}$. [b] $B^2/B_0^2$, i.e., the magnetic energy density normalized to its upstream (initialized) value. The over-plotted white contours are magnetic field lines. [c] Local average $\langle\gamma_{\rm e}^2\beta_{\rm e}^2\rangle = \langle\gamma_{\rm e}^2\rangle - 1$, which is proportional to the IC power per electron (the local average is calculated by averaging over patches of $5\times5$ cells). [d] $\Gamma^2-1$, where $\Gamma$ is the bulk Lorentz factor defined in the text. [e] $\langle\gamma_{\rm e}^2\rangle - \Gamma^2$, which represents internal electron motions.
}
\label{fig:2D_rec_layer}
\end{figure*}

In this section, we present the effect of different magnetizations ($1<\sigmai<3$) on the structure of the radiatively-cooled reconnection region, for a fixed $\gcool=0.06$. As we show in Fig.~\ref{fig:recrate}, the reconnection rate exhibits a very weak dependence on the ion-to-electron mass ratio. On the other hand, $\eta_{\rm rec}$ increases from $\sim$0.08 to $\sim$0.12 as $\sigmai$ is increased from 1 to 3; this is consistent with MHD results of \cite{Ni+12} and PIC results of, e.g., \cite{rowan_17} and \cite{ball_18}. For our fiducial mass ratio ($\mi/\me=29$), Fig.~\ref{fig:recrate} also shows that reconnection proceeds with a rate that is nearly independent on the strength of radiative cooling; this conclusion holds both for the electron-ion composition considered here, as well as for the pair plasma of our previous studies (\citetalias{Sironi_20}, \citetalias{Sridhar+21c}). This can be understood from the fact that the reconnection rate is controlled by the electron inertial and thermal effects at X-points, and the electron dynamics at X-points are not affected by cooling for our choice of $\gcr>\sigmae/8$.

The 2D structure of strongly IC-cooled ($\gcool=0.06$) reconnection layers for different magnetizations is presented in Fig.~\ref{fig:2D_rec_layer} ($\sigmai=1$ on the left, $\sigmai=3$ on the right). Regardless of the strength of IC cooling and the value of the magnetization, the reconnection region breaks into a hierarchical chain of coherent magnetic structures (`plasmoids') as a result of the tearing instability \citep{tajima_shibata_97, loureiro_07, bhattacharjee_09, uzdensky_10}. The formed plasmoids occasionally merge with each other, leading to the formation of secondary reconnection layers at the merger site, transverse to the main layer. 

The plasmoid density structure in trans-relativistic ($\sigma\sim 1$) reconnection (Fig.~\ref{fig:2D_rec_layer}) shows some differences  as compared to the ultra-relativistic ($\sigma\gg1$) case of our earlier works (\citealt{sironi_16}, \citetalias{Sironi_20}, \citetalias{Sridhar+21c}). For high magnetizations, plasmoids were nearly circular, and markedly symmetric around their `O'-point---the location of local maximum in the magnetic vector potential, corresponding to the plasmoid center. In contrast, for $\sigma\sim 1$ they display an asymmetric `teardrop' shape, with elongated tails. Older plasmoids (seen near the edges of the box) are more teardrop-like than newborn plasmoids (near the center of the box). This is because large plasmoids grow by accreting smaller (so, faster) plasmoids from behind. Since for $\sigma\sim 1$ all plasmoids have comparable velocities (see below), the accreted plasmoids take some time to merge into the leading plasmoid, which gives the appearance of a long trailing tail.

The density structure in strongly cooled electron-ion simulations appears different from their pair-plasma counterparts (\citetalias{Sridhar+21c}). Plasmoids formed in pair plasma were devoid of particles near their `heads' (in the direction of their motion), which rather accumulate near the plasmoid `tails'. This was caused by Compton drag acting on the electron-positron pairs. Ions, on the other hand, are not appreciably affected by Compton drag. Electrons are electrostatically coupled to ions, and so they inherit the ions' larger inertia---ions have a larger rest mass, and they do not lose the energy gained from reconnection. Thus, in electron-ion simulations the drag effect on the plasma inside plasmoids is weaker than in otherwise equivalent $e^\pm$ simulations.\footnote{Similar results were observed in the hybrid experiment with cooled electrons and uncooled positrons that we showed in \citetalias{Sridhar+21c}.}

A 2D map of the magnetic energy density (in units of its initialized value) is presented in panels [b] of Fig.~\ref{fig:2D_rec_layer}. The peak values of $B^2/B_0^2$---seen at the core of plasmoids---are smaller for $\sigmai=1$, as compared to $\sigmai=3$. The magnetic field in the plasmoid cores is compressed by a factor of $B/B_{\rm 0}\sim14$ for $\sigmai=3$, whereas $B/B_{\rm 0}\sim8$ for $\sigmai=1$.

The total IC power per electron is shown in panels [c] of Fig.~\ref{fig:2D_rec_layer}. This is represented by the quantity $\langle\gamma_{\rm e}^2\beta_{\rm e}^2\rangle=\langle\gamma_{\rm e}^2\rangle-1$, where the average for each cell is performed in a local patch of $5\times5$ cells. As expected, the total IC power is larger for $\sigmai=3$ than for $\sigmai=1$, just as a consequence of the greater energy per particle. For both $\sigmai=1$ and 3, the IC power is fairly uniformly distributed over the reconnected plasma, with a slight excess seen around the primary X-point near the center of the layer, in the thin layers between merging plasmoids, and at the heads and tails of plasmoids. This is similar to what was seen in radiatively-cooled relativistic reconnection of an $e^\pm$ plasma (\citetalias{Sironi_20}, \citetalias{Sridhar+21c}).

The bulk motion's contribution to the total IC power is shown in panels [d] of Fig.~\ref{fig:2D_rec_layer}. This is represented by $\Gamma^2-1$, where $\Gamma=(1-\beta^2)^{-1/2}$ is the bulk Lorentz factor of the plasma, and $\bmath{\beta}$ is the average particle fluid velocity---weighted with the mass densities of both electrons and ions and averaged, for each cell, in the local patch of neighboring $5\times5$ cells. In the frame moving with velocity $\bmath{\beta}$, the plasma stress-energy tensor has a vanishing energy flux, and we define this to be the plasma comoving frame \citep{rowan_19}. We find that the bulk motions' contribution to the IC power is stronger for $\sigmai=3$ than for $\sigmai=1$. This is expected, since faster bulk motions are associated with higher values of the Alfv\'{e}n speed $v_{\rm A} \propto \sqrt{\sigma/(1+\sigma)}$ (more on this in \S\ref{subsubsec:bulk} and Fig.~\ref{fig:ux}). 

The similarity between panels [c] and [d] in Fig.~\ref{fig:2D_rec_layer} for both $\sigmai=1$ and 3  suggests that most of the IC power comes from bulk motions. This is further supported by panels [e] (the difference between panels [c] and [d]) where we show the contribution of the electrons' internal motion (represented by $\langle\gamma_{\rm e}^2\rangle-\Gamma^2$) to the total IC power. The figure shows that the IC power in internal motions is larger in $\sigmai=3$ than in $\sigmai=1$ simulations; this is expected since $\langle\gammae\rangle\propto\sigmae\propto\sigmai$, for the freshly accelerated electrons. While most of the electrons in the reconnection region are cooled down to non-relativistic temperatures, narrow slivers of hot electrons are seen at X-points, where the energization timescale is faster than the IC cooling timescale (\citetalias{Sironi_20}, \citetalias{Sridhar+21c}). The outskirts of plasmoids also appear hot because the electrons accelerated at X-points of the main layer and at plasmoid merger sites enter into neighboring plasmoids along their outer boundaries. Nonetheless, the internal motions' contribution to the IC power is highly sub-dominant relative to the one of bulk motions, in most of the layer. This conclusion, which we had derived for pair plasmas  (\citetalias{Sironi_20}, \citetalias{Sridhar+21c}), also holds in electron-ion plasmas.

\subsection{Bulk and internal motions} \label{subsec:motions}

In this section, we explore in more detail the effects of plasma magnetization and IC cooling strength on the bulk and internal motions of the reconnected plasma. We define the reconnection region as the area around the reconnection midplane $y = 0$ that contains a mixture of particles coming from above and below it, with both populations contributing at least 1\% to the mixture \citep{rowan_17}. In the following sections, our discussion of the particle motions involve plasmas with different compositions (i.e., electron-ion vs. $e^\pm$ pair). Therefore, the magnetization experienced by electrons should be cast as $\sigmae=\sigma\mi/\me$ for the electron-ion case (where the overall magnetization $\sigma\sim\sigmai$) and as $\sigmae=2\sigma$ for the $e^\pm$ case (where $\sigma=\sigma_{e^\pm}$). We demonstrate below that the  electron bulk motions are primarily determined by the overall $\sigma$, whereas their internal motions are governed by $\sigmae$.

\subsubsection{Bulk component} \label{subsubsec:bulk}

\begin{figure}
\includegraphics[width=8cm]{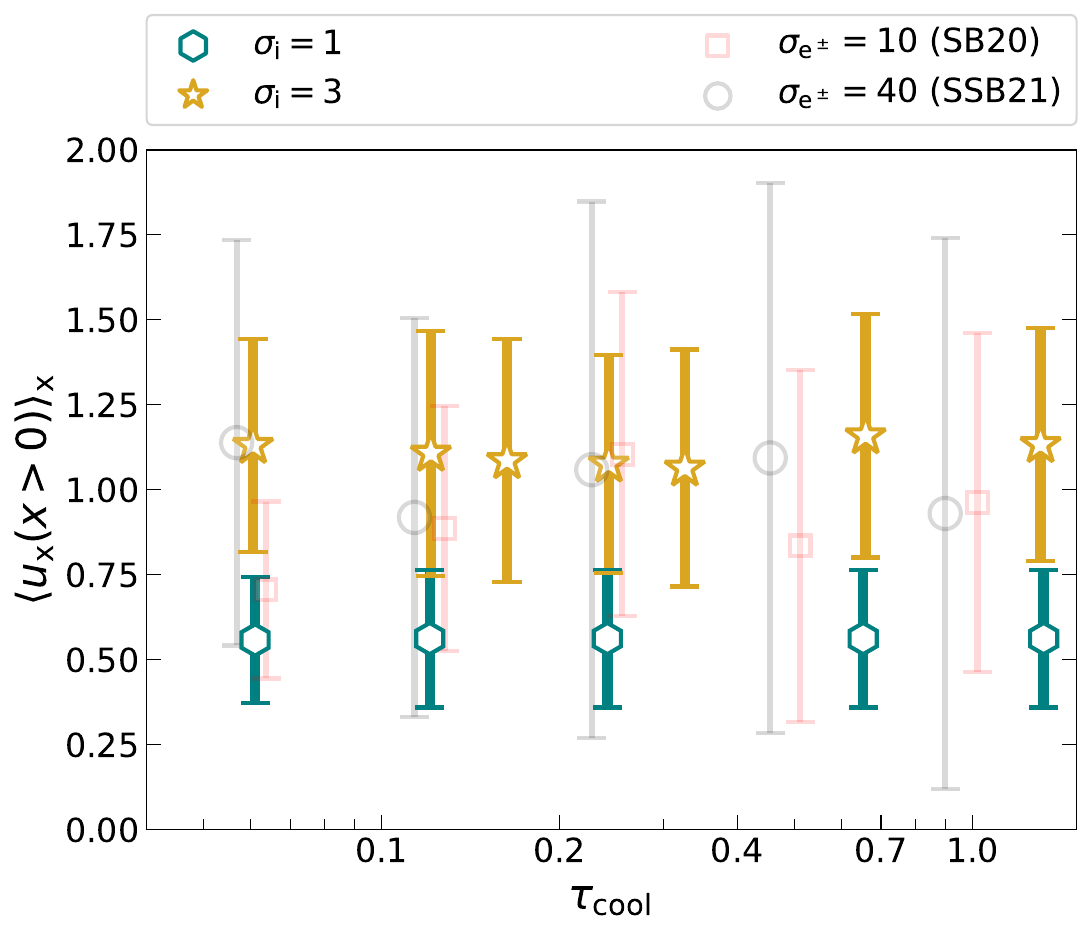}
\caption{Bulk outflow motions in the reconnection region as a function of $\tau_{\rm cool}$ and $\sigmai$. Green hexagons and brown stars indicate the mean (density-weighted) bulk 4-velocity $\ux$ for $\sigmai=1$ and 3 respectively, and the error bars indicate its standard deviation. The red squares and black circles denote the same quantity for an $e^\pm$  plasma with $\sigma_{e^\pm}=10$ (\citetalias{Sironi_20}) and $\sigma_{e^\pm}=40$ (\citetalias{Sridhar+21c}), respectively. The measurements were performed along one half of the reconnection region ($x>0$) and the mean and standard deviation are computed during the time interval $2 \lesssim T/(\Lx/v_{\rm A}) \lesssim 5$.}
\label{fig:ux}
\end{figure}

To quantify the  bulk motions along the outflow direction, we calculate the bulk 4-velocity $\ux=\Gamma\beta_{\rm x}$. To ensure sufficient statistics, we calculate $\ux$ only in those cells with at least 4 particles, and then compute the density-weighted average in space (across the right half of the box with $x/\Lx>0$) and time (in the quasi-steady interval: $2 \lesssim T/(\Lx/v_{\rm A}) \lesssim 5$). This quantity, $\langle \ux(x>0)\rangle_{\rm x}$, is presented in Fig.~\ref{fig:ux} for different $\gcool$ and $\sigmai$. The error bars illustrate the stochasticity of the reconnection layer, i.e., the dispersion away from the mean bulk motion. We quantify this by the density-weighted spatio-temporal average of the standard deviation of $\ux$ (\citetalias{Sridhar+21c}). Such stochastic motions can be crucial in determining the Comptonized spectrum. They can arise due various reasons including the different speeds of plasmoids with different sizes, and the fact that a few small plasmoids can move against the mean flow, being pulled back by a large plasmoid behind them.

Fig.~\ref{fig:ux} demonstrates that for a given $\sigma$ the mean bulk outflow velocity $\langle \ux \rangle$, barely depends upon $\gcool$---akin to strongly cooled $e^\pm$ plasmas with $\sigma=\sigma_{e^\pm}>10$ (\citetalias{Sridhar+21c}). For a given $\gcool$, $\langle \ux \rangle$ varies between $\sim$0.55 for $\sigma\simeq\sigmai=1$ and $\sim$1.1 for $\sigma\simeq\sigmai=3$. On the other hand, $\langle \ux \rangle$ is seen to be similar between electron-ion models with $\sigma=\sigmai=3$, and  pair plasma simulations with $\sigma=\sigma_{e^\pm}=10$ or 40.  This shows that $\langle \ux \rangle$ is $\sigma$-independent, and also independent of the plasma composition, for magnetizations $\sigma \gtrsim3$. 

A systematic increase in the stochasticity of the bulk outflow motions with increasing $\sigma$ was identified in \citetalias{Sridhar+21c} for $e^\pm$ plasmas. We broadly observe a similar effect here for electron-ion plasma. Considering the overall range of magnetizations, while the mean values $\langle\ux\rangle$ for $\sigmai=3$ and $\sigma_{e^\pm}=40$ are nearly the same, the dispersion in the bulk outflow motions monotonically increases with magnetization, from $\sigmai=1$ up to $\sigma_{e^\pm}=40$. This effect can be explained as follows. A higher dispersion is obtained for larger deviations of $\ux$ from the mean value. The largest $\ux$ is typically attained by the smallest plasmoids, and is given by the Alfv\'{e}nic limit of $u_{\rm x,max}\sim\sqrt{\sigma}$ (\citetalias{Sironi_20}, \citetalias{Sridhar+21c}). We find that $u_{\rm x,max}/\langle\ux\rangle\sim 1.8$ for $\sigma=\sigmai=1$, whereas it is $\sim$5.7 for $\sigma=\sigma_{e^\pm}=40$. This result suggests that vigorous collisions of plasmoids (i.e., colliding with appreciably different speeds) are relatively less likely to occur in the low-$\sigma$ regime considered here, as compared to the higher magnetization of our earlier works (\citetalias{Sironi_20}, \citetalias{Sridhar+21c}). As an additional consequence, the secondary transverse reconnection layers (along $y$) formed between colliding plasmoids are rarer at lower $\sigma$. Therefore, the  contribution of $y$-directed bulk motions to the Comptonized emission gets smaller with lower $\sigma$.

\subsubsection{Internal component} \label{subsubsec:internal}

\begin{figure}
  \includegraphics[width=8cm]{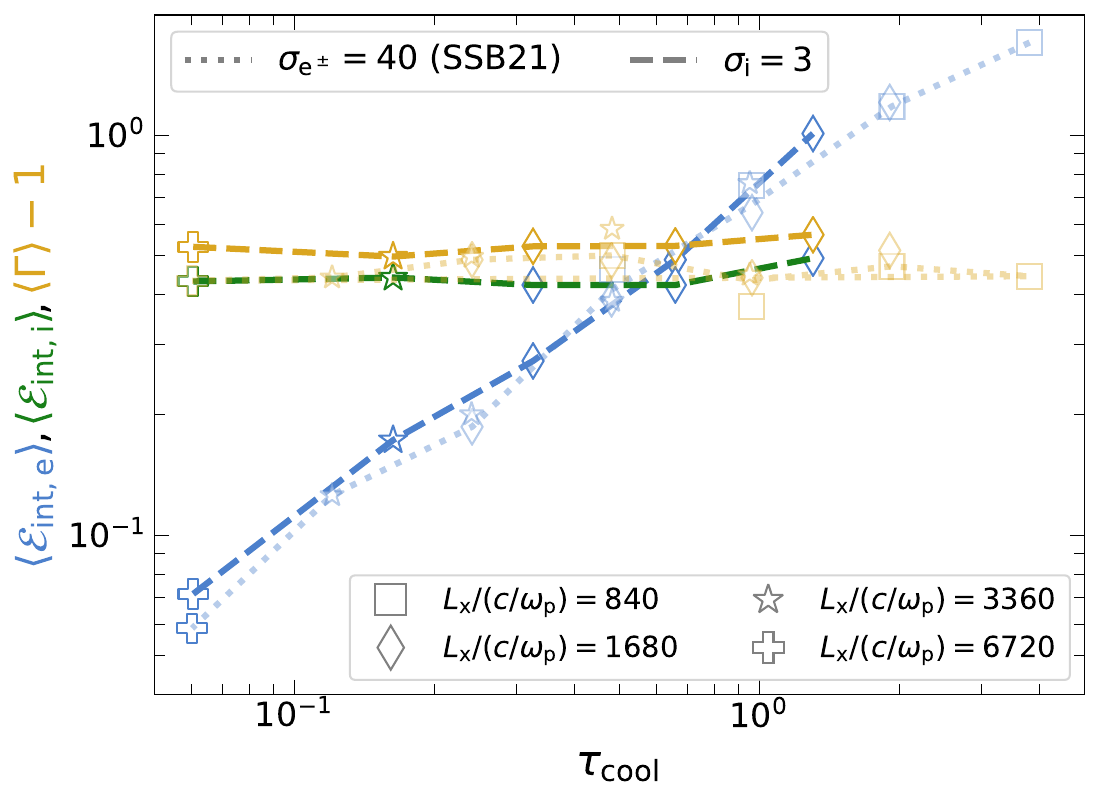}  
  \caption{Average particle energy (in units of $m_{\rm s}c^2$) separated into two components: internal (blue for electrons and green for ions) and bulk (brown), plotted against $\gcool$. Dashed curves refer to $\sigmai=3$, and dotted lines to $\sigma_{\rm e^\pm}=40$ (\citetalias{Sridhar+21c}). For a given magnetization and $\gcool$, different symbols correspond to different sizes of the simulation domain (see legend), which equivalently correspond to different $\gcr$. Where more than one simulation is available with the same $\sigma$ and $\gcool$ (Table~\ref{tab:setup}), the curve passes through their mean value. The measurements were performed using density-weighted averaging over the reconnection region and averaging over the quasi-steady-state time interval $2 \lesssim T/(\Lx/v_{\rm A}) \lesssim 5$.}
\label{fig:energies_taucool_sigma}
\end{figure}

In this section, we set out to investigate how the energy in the bulk motion compares against the internal energy (heat) of electrons and ions, for a wide range of IC cooling strengths ($0.06\lesssim\gcool\lesssim3$). We specifically choose to compare the $\sigma=\sigmai=3$ and $\sigma=\sigma_{e^\pm}=40$ models because they both correspond to the same electron magnetization $\sigmae\sim80$ (for our choice of $\mi/\me=29$ in electron-ion simulations). Fig.~\ref{fig:energies_taucool_sigma} shows the mean bulk energy per particle ($\langle\Gamma\rangle-1$) and the mean internal energy per particle ($\varepsilon_{\rm int,e}$ for electrons, $\varepsilon_{\rm int,i}$ for ions, normalized to their respective rest mass energies)\footnote{We refer readers to Appendix~C in our Paper I (\citetalias{Sridhar+21c}) for a detailed discussion of how the internal energies are calculated.}. The mean values are obtained by first performing a density-weighted average of the respective quantity over the reconnection region, and then a time-average over the interval $2 \lesssim T/(\Lx/v_{\rm A}) \lesssim 5$.

Fig.~\ref{fig:energies_taucool_sigma} demonstrates that the mean bulk energy ($\langle\Gamma\rangle-1\sim0.5$) is independent of $\gcool$ and $\sigma$ for sufficiently relativistic plasmas, $\sigma\gtrsim3$ (see also Fig.~\ref{fig:ux}). As far as the internal energy is concerned, our results reveal a remarkable similarity in $\langle \varepsilon_{\rm int,e}\rangle$ between electron-positron models with $\sigma=\sigma_{e^\pm}=40$ and electron-ion cases with $\sigmai=3$, across a wide range of $\gcool$. This is because the electron internal motions are determined by the electron magnetization $\sigmae$, which controls the mean energy available per electron, and is similar between the $\sigma=\sigma_{e^\pm}=40$ ($\sigmae=80$) and $\sigma=\sigmai=3$ ($\sigmae=85.5$) models, for $\mi/\me=29$. We see that $\langle\varepsilon_{\rm int,e}\rangle > \langle\Gamma\rangle-1$ in the `hot regime' ($\gcool\gtrsim0.5$), whereas in the `cold regime' ($\gcool\ll1$) electrons are cooled down to non-relativistic temperatures, so their total energy is dominated by bulk motions. 

In our electron-ion plasma simulations with $\sigma=\sigmai=3$, the ion internal energy $\langle\varepsilon_{\rm int,i}\rangle$ stays at trans-relativistic values $\sim0.4$ regardless of $\gcool$. The fact that $\langle\varepsilon_{\rm int,i}\rangle$ does not depend on $\gcool$ allows one to conclude that 
energy is not transferred from the ions to the electrons, even when electrons are strongly cooled. In general, the electrons and ions are thermally decoupled if both collisional (Coulomb) and collisionless couplings are inefficient. Coulomb coupling is expected to be inefficient in the coronal plasma with $\sigma>1$ (Appendix~\ref{appendix:thermalization}; there, we also show that our simulations are well within the collisionless regime). Collisionless channels for ion-to-electron energy transfer could in principle occur, however we find they are not activated. 
As we describe in Appendix~\ref{appendix:temp_anisotropy}, we have directly verified that the layer is not prone to collisionless ion velocity-space instabilities (e.g., ion-cyclotron, mirror, and firehose, see \citealt{sironi_15}). 

Also note in Fig.~\ref{fig:energies_taucool_sigma} that the value and trend of $\langle\varepsilon_{\rm int,i}\rangle$ as a function of $\gcool$ is identical to that of $\langle\Gamma\rangle-1$. This is generally expected in mildly-relativistic reconnection ($\sigma\sim1$): here, not only does the internal energy of ions scale with $\sigmai$, so also does their bulk energy, $\langle\Gamma\rangle-1\sim \langle\beta^2\rangle/2\sim\sigma\sim\sigmai$ (neglecting factors of order unity).

\subsection{Particle energy spectra} \label{subsec:spectra}

\begin{figure*}
\includegraphics[width=16cm]{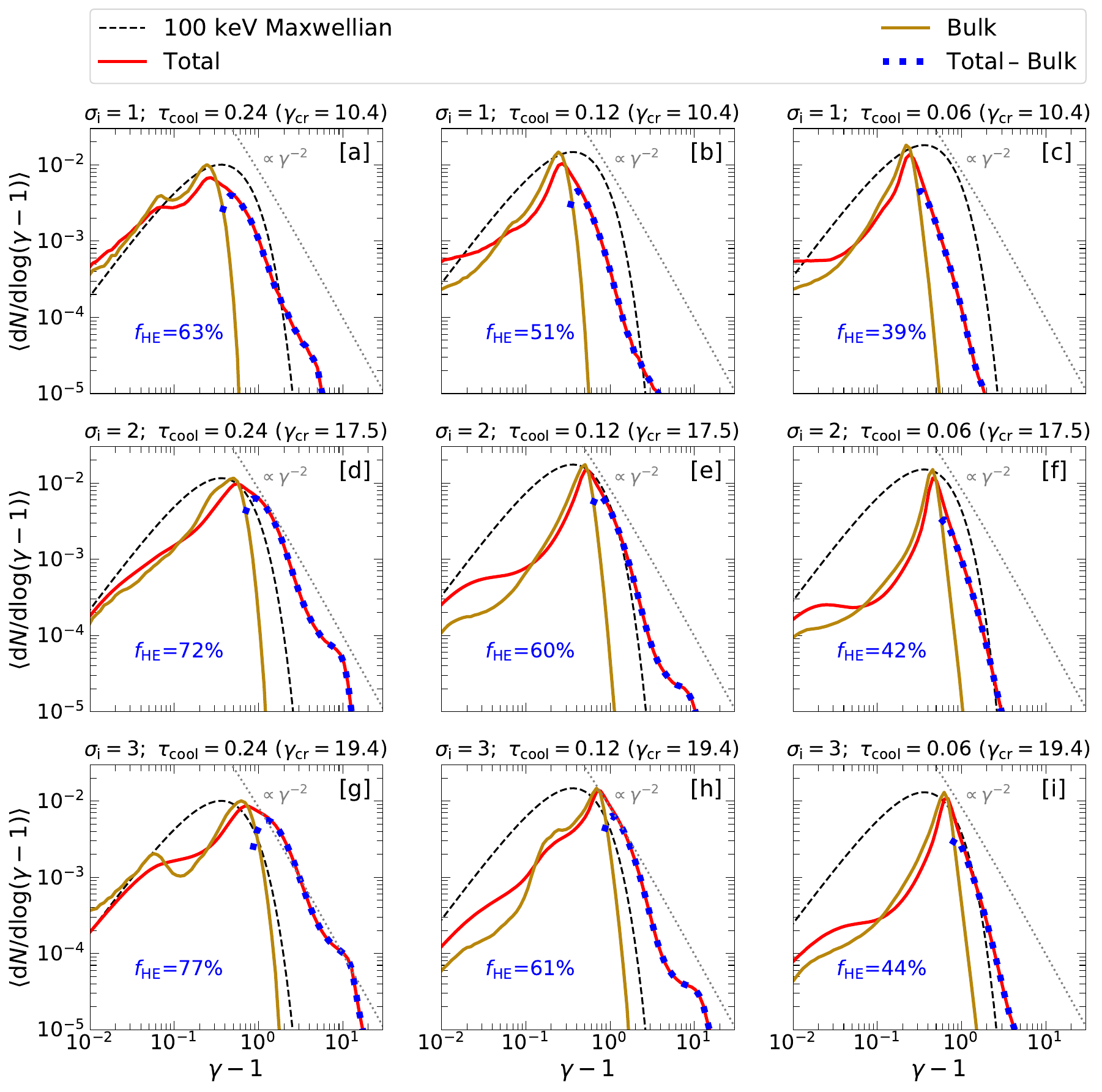}
\caption{Electron energy spectra extracted from the reconnection region and averaged over $2 \lesssim T/(\Lx/v_{\rm A})\lesssim 5$. Nine simulations are shown with magnetizations $\sigmai=1,2,3$ and different values of $\gcool$ and domain size $\Lx$. The plasma magnetization increases from top to bottom, and the effective cooling strength (parametrized by $\gcool$) increases from left to right (achieved by increasing $\Lx$ at fixed $\gcr>\sigmae/8=\sigmai(\mi/8\me)$). The spectra are normalized by $\Lx^2$ for proper comparison of models with different $\Lx$. Each panel shows the total energy spectrum (red; $\gamma=\gammae$), the bulk motion spectrum (golden-brown; $\gamma=\Gamma$), and their difference (dotted blue), i.e. the part not accounted for by bulk motions; the fractional IC power contributed by this component is denoted in each panel as $f_{\rm HE}$. For comparison, we also plot a Maxwellian distribution with a temperature of 100\,keV (dashed black), such that the normalization of its peak matches the peak of the bulk energy spectrum. The dotted grey lines indicate the $-2$ slope that corresponds to equal IC power per decade in Lorentz factor. 
}
\label{fig:spectra_sig123_cooling}
\end{figure*}

Here, we present the electron and ion energy spectra for particles belonging to the reconnection region. The spectra are averaged in time over $2 \lesssim T/(\Lx/c) \lesssim 5$---representative of a quasi-steady state. Since $\gcool\propto\gcr^2/\Lx$, we can in principle modify either $\gcr$ or the size of the computational domain $\Lx$ (or both), in order to vary $\gcool$. For a given $\sigma$, we choose to fix $\gcr$ to a value such that $\gcr>\sigmae/8$ (see \S\ref{sec:IC_cooling}), and only modify $\Lx$. The spectra are then normalized by $\Lx^2$ for a fair comparison between simulations with different box sizes.

\begin{figure}
\includegraphics[width=8.2cm]{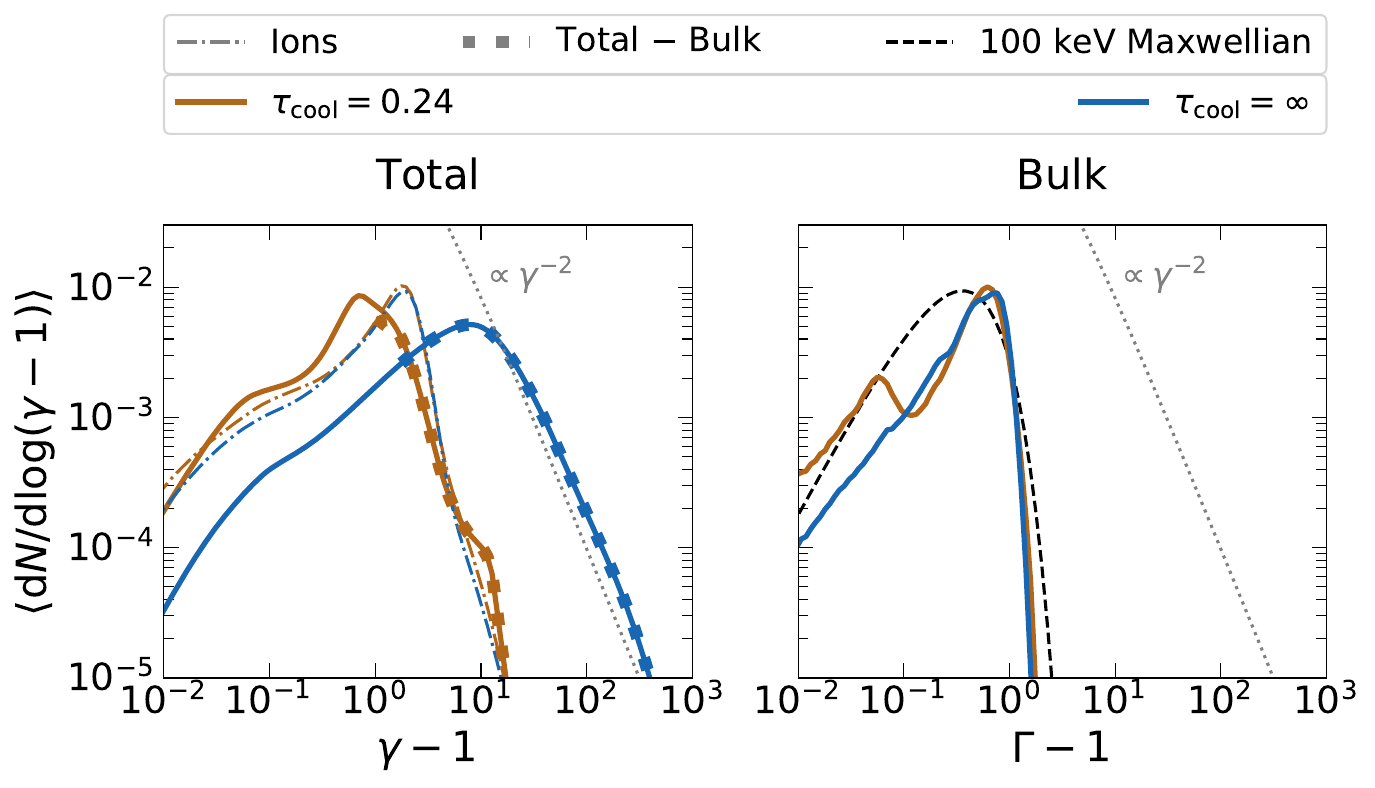}
\caption{Total (left) and bulk (right) energy spectra for a model with $\sigmai=3$ and $\mi/\me=29$. In both  panels, the blue curves denote the uncooled simulations ($\gcool=\infty$), whereas the brown curves denote the cooled cases ($\gcool\simeq0.24$; same case as Fig.~\ref{fig:spectra_sig123_cooling}[g]). The solid curves represent electron spectra whereas the dash-dotted curves denote the ion spectra. For comparison, we also plot a Maxwellian distribution with a temperature of 100\,keV (dashed black; right panel) normalized so that its peak matches the peak of the bulk energy spectrum. The dotted grey lines indicate the $-2$ slope that corresponds to equal IC power per decade in Lorentz factor. The spectra are extracted from the reconnection region of a simulation with size $\Lx/(\compe)=1680$, and time-averaged in the interval $2 \lesssim T/(\Lx/v_{\rm A}) \lesssim 5$.}
\label{fig:spectra_coolnocool}
\end{figure}

Fig.~\ref{fig:spectra_sig123_cooling} presents the total energy spectra of electrons for different models: the magnetization increases from $\sigmai=1$ to 3 from top to bottom row, and the strength of IC cooling increases from $\gcool=0.24$ to 0.06 from left to right column. In addition to the total energy spectrum, each panel also shows the two chief components of the total spectrum---the bulk motion, and the high-energy tail dominated by the internal motion. For completeness, we also compare  in Fig.~\ref{fig:spectra_coolnocool} the spectra of electrons and ions, between uncooled ($\gcool=\infty$) and cooled cases ($\gcool=0.24$) for $\sigmai=3$. Note that the cooled model is the same as the one in Fig.~\ref{fig:spectra_sig123_cooling}[g].

In the absence of IC cooling, the total energy spectrum (left panel of Fig.~\ref{fig:spectra_coolnocool}) of electrons in a $\sigmai=3$ ($\sigmae\sim85$) electron-ion plasma peaks at $\gammae-1\sim10$, which resembles that of the $\sigma_{e^\pm}=40$ ($\sigmae=80$) model in $e^\pm$ plasma (\citetalias{Sridhar+21c}). This is because the mean Lorentz factor of the electrons energized by reconnection is proportional to $\sigmae$, and in the absence of IC cooling, electrons retain the energy they acquired from field dissipation. On the other hand, in the presence of significant IC cooling ($\gcool=0.24$), the peak in the electron total spectrum shifts to trans-relativistic energies, $\gammae-1\sim0.7$ for the $\sigmai=3$ model. By comparison with the right panel of Fig.~\ref{fig:spectra_coolnocool}, we conclude that in this regime the electron energy is controlled by bulk motions. In contrast, the total energy spectra of ions are insensitive to the strength of IC cooling, and peak at $\gamma_{\rm i}-1\sim2$ for $\sigmai=3$.

Fig.~\ref{fig:spectra_sig123_cooling} shows that, in the strong cooling regime ($0.24>\gcool>0.06$), the bulk component of the spectra peaks stably at trans-relativistic speeds, with a moderate increase with magnetization: $\Gamma-1\sim0.2$, 0.4 and 0.7, for $\sigmai=1$, 2 and 3, respectively. These are the energies where the total energy spectra peak as well, indicating that in the regime of strong cooling the electron energy is dominated by the bulk motion (see also \S\ref{subsec:motions} and Fig.~\ref{fig:energies_taucool_sigma}). This conclusion, which is especially clear for models with $\gcool=0.06$, is in agreement with our earlier results for $e^\pm$ plasma (\citetalias{Sironi_20}, \citetalias{Sridhar+21c}).

For all the models presented in Fig.~\ref{fig:spectra_sig123_cooling}, we also plot a Maxwell-J\"{u}ttner distribution with $k_{\rm B}T_{\rm e}=100$\,keV. We showed in our earlier works (\citetalias{Sironi_20}, \citetalias{Sridhar+21c}) that the spectrum of bulk motions in IC-cooled $\sigma\gg1$  pair plasmas resembles a thermal distribution with $k_{\rm B}T_{\rm e}=100$\,keV. The bulk motion spectra in our electron-ion plasmas---for $1\lesssim\sigmai\lesssim2$---also have peak energies comparable to the same Maxwellian. Yet, the bulk spectra for our electron-ion cases are narrower than the Maxwellian, whereas they had comparable widths in our earlier electron-positron studies (\citetalias{Sridhar+21c}). This 
can be understood from the level of stochasticity in the plasma outflow. In fact, the reconnection region is more disordered---with plasmoids moving at a wide range of speeds---for larger $\sigma$ (see also Fig.~\ref{fig:ux}). Such stochastic motions will manifest as a broader energy spectrum---with a wider peak---for higher magnetizations.

The higher energy tail of the spectra (i.e., blue dotted curves in Fig.~\ref{fig:spectra_sig123_cooling}), cannot be accounted for by the bulk motions alone, and is instead dominated by the internal motions. As shown in Fig.~\ref{fig:2D_rec_layer}, the electrons belonging to this component are produced at inter-plasmoid regions (near X-points) in the main layer or at the interface of merging plasmoids (we refer to \citetalias{Sironi_20} and \citetalias{Sridhar+21c} for more details). These electrons do radiate a considerable fraction ($\fHE$) of the total IC power, which is quoted in each panel of Fig.~\ref{fig:spectra_sig123_cooling}. For our considered models, $\fHE$ ranges from 39\% (for $\sigmai=1$, $\gcool=0.06$) to 77\% (for $\sigmai=3$, $\gcool=0.24$). For a given $\gcool$, $\fHE$ increases by $\sim9\%$ from $\sigmai=1$ to 3 and, for a given $\sigmai$, $\fHE$ decreases by $\sim28\%$ from $\gcool=0.24$ to 0.06. We also find that, for a given $\gcool$, $\fHE$ is generally larger in electron-ion simulations (for $\sigmai\geq1$) than it is in pair plasma simulations (for $\sigma_{e^\pm}\gg1$; \citetalias{Sridhar+21c}). This is because the high energy component of the spectrum that determines $\fHE$ is defined as the difference between the total spectrum and its bulk component; the narrower shape of the bulk spectrum in our electron-ion simulations---compared to our pair plasma cases---then enhances the impact of the high energy component, thus leading to a larger $\fHE$.

The high energy component displays a `knee' at $\gammae-1\sim10$. This is populated predominantly by electrons accelerated at strong X-points near the center of the reconnection layer. The extent over which such X-points appear is $\sim 500\,c/\omega_{\rm pe}$ regardless of $L_{\rm x}$, which implies that the prominence of the high-energy knee decreases with increasing box size, since the fractional importance of the central X-points drops. This can be seen from all rows of Fig.~\ref{fig:spectra_sig123_cooling} where the simulation box size increases from left to right. Even larger simulations will be needed to assess whether the normalization of the high-energy knee eventually settles to a fixed (i.e., box-independent) fraction of the overall particle count, or whether it keeps dropping as $\propto L_{\rm x}^{-1}$.

\section{Monte Carlo radiative transfer calculations} \label{sec:radiative_transfer}

\begin{figure*} 
\includegraphics[width=13cm]{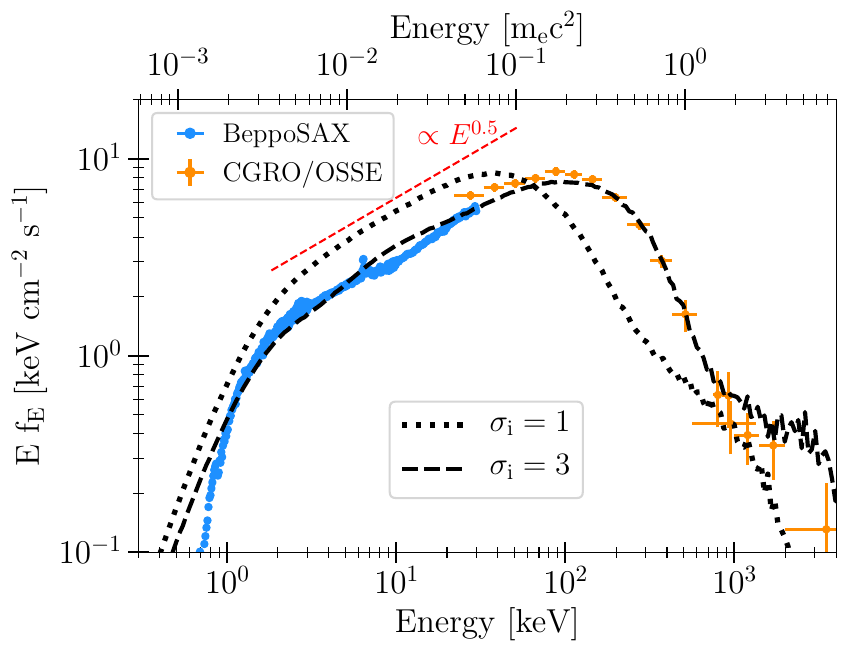}
\caption{X-ray/$\gamma$-ray spectrum of Cygnus~X-1 during its hard state, $Ef_{\rm E}=E^2N(E)$. The 0.7--25\,keV data (blue) are from \textit{BeppoSAX} \citep{DiSalvo+01, Frontera+01}, and the 25--3500\,keV data (orange) are from \textit{CGRO/OSSE} (\citealt{McConnell+02}; see \citealt{Zdziarski+17} for the spectrum in an extended range). The black dotted and dashed curves show the spectra formed by Comptonization in the reconnection layer in the models with $\sigmai=1$ (dotted) and $\sigmai=3$ (dashed), with the same strength of radiative cooling losses ($\gcool=0.06$) and Compton amplification factor ($A=10$). The dashed red line indicates the power-law $N(E)\propto E^{-1.5}$. All the data are normalized with respect to \textit{OSSE}.}
\label{fig:X-ray_spectra}
\end{figure*}

The X-ray emission expected from the reconnection layer is calculated by performing Monte Carlo simulations of radiative transfer, using the code \texttt{CompPair} (\citet{Beloborodov_20}, \citet{Sridhar+21c}). In the Monte Carlo simulation, we inject soft seed photons sampled from a Planck distribution of temperature $k_{\rm B}T_{\rm s}=10^{-3}m_{\rm e}c^2$ into the mid-plane of the reconnection layer. The photons are then followed in time, as they scatter across the reconnection region, thereby producing a Comptonized, hard X-ray spectrum. 

We divide the region $|x|<\Lx$, $|y|<H=\Lx/3$ into $60\times 20$ patches, and for each photon scattering event, the particle momentum is drawn from the local distribution function in the patch. The transfer simulation is then evolved in time (between $T/(\Lx/v_{\rm A})=2.5$ and $T/(\Lx/v_{\rm A})=5.5$), with updating the particle distribution function every $0.5\Lx/v_{\rm A}$. The overall normalization of the plasma density is parameterized using the Thomson optical depth $\tau_{\rm T}=H\sigma_{\rm T}n_{\rm 0}$. We adjust the optical depth across the reconnection layer to obtain a Compton amplification factor\footnote{The factor $A$ is the ratio of the average energy of escaping photons to the average energy of injected soft photons.} $A\approx 10$, which gives the Comptonized spectrum with a photon index of $\Gamma\sim1.5$ (defined as $dN/dE \propto E^{-\Gamma}$, where $N$ is the number of photons with energy $E$) typical for the hard state spectra of accreting black holes. In analogy to the PIC simulation, we employ outflow boundary conditions for the Monte Carlo calculation (see Appendix~\ref{appendix:Monte_Carlo_bc} for different choices of boundary conditions).

Fig.~\ref{fig:X-ray_spectra} shows the spectrum of radiation escaping from the reconnection region for two models, with $\sigmai=1$ (dotted) and $\sigmai=3$ (dashed), adopting the same $\tau_{\rm cool}=0.06$. The emitted spectrum varies with time as reconnection proceeds, and Fig.~\ref{fig:X-ray_spectra} shows the emission averaged over the time interval $2.5\lesssim T/(L_{\rm x}/c)\lesssim5.5$. We observe that the spectrum peaks around 40\,keV for $\sigmai=1$ and 100\,keV for $\sigmai=3$. This shift is primarily caused by the increase in plasmoid speeds at higher $\sigmai$, see Fig.~\ref{fig:ux}. The model with $\sigmai=3$ (dashed) is remarkably close to the typical hard-state of Cygnus~X-1 observed in a broad band of $1-1000$\,keV. For comparison, we over-plot in Fig.~\ref{fig:X-ray_spectra} the data available from \textit{BeppoSAX} and \textit{CGRO/OSSE}. The deviation of the data from the model at low photon energies ($\lesssim 1$\,keV) is expected from inter-stellar absorption. The observed excess at higher energies ($\gtrsim$1\,MeV) is also well reproduced by our $\sigmai=3$ model. However, we caution that the particles responsible for this high-energy component belong to the high-energy knee in Fig.~\ref{fig:spectra_sig123_cooling}, whose prominence may become vanishingly small in the asymptotic (realistic) limit of very large domains. In this case, the high-energy photons  at $\gtrsim$1\,MeV may be due to coronal regions with $\sigmai>3$.

\section{Summary}  \label{sec:summary}
The main goal of this series of papers is to explore the mechanism that gives rise to the observed hard X-ray spectra of accreting black holes. In the previous papers (\citetalias{Sironi_20}, \citetalias{Sridhar+21c}), we performed 2D and 3D PIC simulations of relativistic magnetic reconnection in $e^\pm$ plasmas subject to strong IC cooling---as appropriate to the most magnetically dominated regions of black hole coronae closer to the jet---and found that thermal Comptonization of soft seed photons by hot electrons is an unfeasible mechanism. In fact, due to the strong IC losses, electrons get cooled down to non-relativistic temperatures. Nonetheless, we demonstrated that the hard X-ray spectra can be produced via Comptonization by the bulk motions of the Compton-cooled plasmoid chain \citep{belo_17}. In this paper II, we have performed the first simulations of trans-relativistic magnetic reconnection ($\sigma\sim1$) in electron-ion plasma subject to the effects of IC cooling. This is a regime that can be most relevant to the moderately-magnetized regions of coronae adjacent to the accretion disk. Our main results are as follows.

\begin{enumerate}
\item Regardless of the strength of IC cooling ($\gcool$) or magnetization ($\sigma$), the reconnection layer breaks into a hierarchical chain of plasmoids. We also demonstrate that the rate at which reconnection proceeds is independent of the strength of IC cooling, and nearly independent of the ion-to-electron mass ratio, $\mi/\me$.

\item The bulk motions of the plasmoid chain dominate the total IC power in the strong cooling regime. The mean energy in bulk motions is independent of the strength of IC cooling. On the other hand, the mean 4-velocity of bulk motions increases by a factor of two between $\sigmai=1$ and 3, before saturating at a value that is consistent with our earlier results for larger magnetizations $\sigma\ge10$  (\citetalias{Sridhar+21c}). This shows that the typical speed of bulk motions becomes $\sigma$-independent only for $\sigma\gtrsim3$.

\item For a wide range of IC cooling strengths ($0.06\le\gcool\le1.32$), the internal energy per electron for $\sigmai=3$ behaves remarkably similar to its $\sigma_{e^\pm}=40$ counterpart in $e^\pm$ plasma. This is because the electron magnetization is $\sigmae\sim80$ in both cases, and $\sigmae$ determines the electron internal energy. The electron internal energy scales linearly with $\gcool$ in the range we explored, and attains non-relativistic temperatures $\sim5$\,keV for strong cooling, $\gcool\sim0.06$.

\item The radiatively inefficient ions retain the heat gained from reconnection, and their mean energy stays independent of $\gcool$. Thus, the IC-cooled electrons are not re-heated by the hot ions. This renders the electrons incapable of powering thermal Comptonization, regardless of the plasma composition.

\item The peak of the electrons' bulk energy spectra for IC-cooled trans-relativistic reconnection is generally narrower---indicating a more ordered outflow---than it is for relativistic ($\sigma\gg1$) reconnection in $e^\pm$ plasma. For strong cooling, the total electron spectra are dominated by their bulk component---whose peak energy increases with $\sigmai$. For $\sigmai=2$ the spectral peak coincides with that of a Maxwellian distribution with a temperature of 100\,keV. The high energy tail of the spectra---that are not accounted for by bulk motions---are due to electrons accelerated at X-points, on a timescale faster than IC losses. In the limit of strong cooling, the fraction of radiated IC power contributed by these hot electrons is about $40\%$ for $1\le\sigmai\le3$.

\item We find that reconnection layers in  BH coronae with $\sigmai=3$ can Compton-upscatter soft photons to non-thermal energies and explain the typical hard-state spectrum of accreting black holes. Such a spectrum peaks near 100\,keV, and is primarily shaped by the bulk motions of plasmoids laden with cold (IC-cooled) electrons (whereas ions remain hot). The high-energy MeV tail may be produced by hot electrons quickly energized at X-points.

\end{enumerate}

In the future, it may be useful to investigate how the results change for an electron-positron-ion plasma \citep[e.g.,][for the non-radiative case]{petropoulou_19}, in the presence of a stronger guide field, and for an asymmetric reconnection layer. Also, it has recently been shown that the physics of particle acceleration in reconnection is significantly different between 2D and 3D \citep{zhang_21}. In 2D, electrons are pre-accelerated near X-points up to $\gamma_{\rm X}\sim \sigma_{\rm e}/8$ \citep[e.g.,][]{sironi_22}, and then they get advected into plasmoids, where further energization proceeds slowly \citep[e.g.,][]{petropoulou_18,hakobyan_20}.  Instead, in 3D some of the particles with $\gamma\gtrsim\gamma_{\rm X}$ can escape from plasmoids and experience the large-scale electric field in the inflow region, resulting in fast acceleration \citep{zhang_21}, potentially up to $\gammacr$. Further work is needed to clarify whether the amount of reconnection power dissipated in these high-energy particles  can  change the properties of plasmoid-mediated cold-chain Comptonization.

\section*{Acknowledgements}

This paper benefited from useful discussions with Riley M.~T.~Connors, Javier A.~Garc\'{i}a, Victoria Grinberg, Guglielmo Mastroserio, James F.~Steiner, and Aaron Tran. We thank Andrzej Zdziarski for sharing details pertaining to the hard state observations of Cygnus~X-1. N.S. acknowledges the support from NASA 80NSSC22K0332. L.S. acknowledges support from the Cottrell Scholars Award, NASA 80NSSC20K1556, NSF PHY-1903412, DoE DE-SC0021254 and NSF AST-2108201. A.M.B. is supported by grants from NSF AST-1816484 and AST-2009453, NASA 21-ATP21-0056, Simons Foundation \#446228, and the Humboldt Foundation. This project made use of the following computational resources: NASA Pleiades supercomputer, Cori of National Energy Research Scientific Computing Center and, Habanero and Terremoto HPC clusters at Columbia University.

\section*{Data availability}
The data underlying this article will be shared on reasonable request to the authors.

%%%%%%%%%%%%%%%%%%%%%%%%%%%%%%%%%%%%%%%%%%%%%%%%%%


\bibliographystyle{mnras}
\bibliography{main} % if your bibtex file is called example.bib

\begin{thebibliography}{}
\makeatletter
\relax
\def\mn@urlcharsother{\let\do\@makeother \do\$\do\&\do\#\do\^\do\_\do\%\do\~}
\def\mn@doi{\begingroup\mn@urlcharsother \@ifnextchar [ {\mn@doi@}
  {\mn@doi@[]}}
\def\mn@doi@[#1]#2{\def\@tempa{#1}\ifx\@tempa\@empty \href
  {http://dx.doi.org/#2} {doi:#2}\else \href {http://dx.doi.org/#2} {#1}\fi
  \endgroup}
\def\mn@eprint#1#2{\mn@eprint@#1:#2::\@nil}
\def\mn@eprint@arXiv#1{\href {http://arxiv.org/abs/#1} {{\tt arXiv:#1}}}
\def\mn@eprint@dblp#1{\href {http://dblp.uni-trier.de/rec/bibtex/#1.xml}
  {dblp:#1}}
\def\mn@eprint@#1:#2:#3:#4\@nil{\def\@tempa {#1}\def\@tempb {#2}\def\@tempc
  {#3}\ifx \@tempc \@empty \let \@tempc \@tempb \let \@tempb \@tempa \fi \ifx
  \@tempb \@empty \def\@tempb {arXiv}\fi \@ifundefined
  {mn@eprint@\@tempb}{\@tempb:\@tempc}{\expandafter \expandafter \csname
  mn@eprint@\@tempb\endcsname \expandafter{\@tempc}}}

\bibitem[\protect\citeauthoryear{{Bale}, {Kasper}, {Howes}, {Quataert}, {Salem}
   \& {Sundkvist}}{{Bale} et~al.}{2009}]{Bale+09}
{Bale} S.~D.,  {Kasper} J.~C.,  {Howes} G.~G.,  {Quataert} E.,  {Salem} C.,
  {Sundkvist} D.,  2009, \mn@doi [\prl] {10.1103/PhysRevLett.103.211101}, \href
  {https://ui.adsabs.harvard.edu/abs/2009PhRvL.103u1101B} {103, 211101}

\bibitem[\protect\citeauthoryear{{Ball}, {Sironi}  \& {{\"O}zel}}{{Ball}
  et~al.}{2018}]{ball_18}
{Ball} D.,  {Sironi} L.,   {{\"O}zel} F.,  2018, \mn@doi [\apj]
  {10.3847/1538-4357/aac820}, \href
  {http://adsabs.harvard.edu/abs/2018ApJ...862...80B} {862, 80}

\bibitem[\protect\citeauthoryear{{Ball}, {Sironi}  \& {{\"O}zel}}{{Ball}
  et~al.}{2019}]{Ball+19}
{Ball} D.,  {Sironi} L.,   {{\"O}zel} F.,  2019, \mn@doi [\apj]
  {10.3847/1538-4357/ab3f2e}, \href
  {https://ui.adsabs.harvard.edu/abs/2019ApJ...884...57B} {884, 57}

\bibitem[\protect\citeauthoryear{{Beloborodov}}{{Beloborodov}}{2017}]{belo_17}
{Beloborodov} A.~M.,  2017, \mn@doi [\apj] {10.3847/1538-4357/aa8f4f}, \href
  {http://adsabs.harvard.edu/abs/2017ApJ...850..141B} {850, 141}

\bibitem[\protect\citeauthoryear{{Beloborodov}}{{Beloborodov}}{2020}]{Beloborodov_20}
{Beloborodov} A.~M.,  2020, arXiv e-prints, \href
  {https://ui.adsabs.harvard.edu/abs/2020arXiv201107310B} {p. arXiv:2011.07310}

\bibitem[\protect\citeauthoryear{{Belyaev}}{{Belyaev}}{2015}]{belyaev_15}
{Belyaev} M.~A.,  2015, \mn@doi [\na] {10.1016/j.newast.2014.09.006}, \href
  {http://adsabs.harvard.edu/abs/2015NewA...36...37B} {36, 37}

\bibitem[\protect\citeauthoryear{{Bhattacharjee}, {Huang}, {Yang}  \&
  {Rogers}}{{Bhattacharjee} et~al.}{2009}]{bhattacharjee_09}
{Bhattacharjee} A.,  {Huang} Y.-M.,  {Yang} H.,   {Rogers} B.,  2009, \mn@doi
  [Physics of Plasmas] {10.1063/1.3264103}, \href
  {https://ui.adsabs.harvard.edu/abs/2009PhPl...16k2102B} {16, 112102}

\bibitem[\protect\citeauthoryear{{Birdsall} \& {Langdon}}{{Birdsall} \&
  {Langdon}}{1991}]{birdsall_91}
{Birdsall} C.~K.,  {Langdon} A.~B.,  1991, {Plasma Physics via Computer
  Simulation}

\bibitem[\protect\citeauthoryear{{Buneman}}{{Buneman}}{1993}]{buneman_93}
{Buneman} O.,  1993, {in ``Computer Space Plasma Physics'', Terra Scientific,
  Tokyo, 67}

\bibitem[\protect\citeauthoryear{{Cao}, {Lucchini}, {Markoff}, {Connors}  \&
  {Grinberg}}{{Cao} et~al.}{2021}]{Cao+21}
{Cao} Z.,  {Lucchini} M.,  {Markoff} S.,  {Connors} R. M.~T.,   {Grinberg} V.,
  2021, arXiv e-prints, \href
  {https://ui.adsabs.harvard.edu/abs/2021arXiv211010547C} {p. arXiv:2110.10547}

\bibitem[\protect\citeauthoryear{{Cassak}, {Liu}  \& {Shay}}{{Cassak}
  et~al.}{2017}]{Cassak+17}
{Cassak} P.~A.,  {Liu} Y.~H.,   {Shay} M.~A.,  2017, \mn@doi [Journal of Plasma
  Physics] {10.1017/S0022377817000666}, \href
  {https://ui.adsabs.harvard.edu/abs/2017JPlPh..83e7101C} {83, 715830501}

\bibitem[\protect\citeauthoryear{{Cerutti}, {Philippov}, {Parfrey}  \&
  {Spitkovsky}}{{Cerutti} et~al.}{2015}]{cerutti_15}
{Cerutti} B.,  {Philippov} A.,  {Parfrey} K.,   {Spitkovsky} A.,  2015, \mn@doi
  [\mnras] {10.1093/mnras/stv042}, \href
  {http://adsabs.harvard.edu/abs/2015MNRAS.448..606C} {448, 606}

\bibitem[\protect\citeauthoryear{{Chatterjee}, {Liska}, {Tchekhovskoy}  \&
  {Markoff}}{{Chatterjee} et~al.}{2019}]{chatterjee_19}
{Chatterjee} K.,  {Liska} M.,  {Tchekhovskoy} A.,   {Markoff} S.~B.,  2019,
  \mn@doi [\mnras] {10.1093/mnras/stz2626}, \href
  {https://ui.adsabs.harvard.edu/abs/2019MNRAS.490.2200C} {490, 2200}

\bibitem[\protect\citeauthoryear{{Daughton}, {Scudder}  \&
  {Karimabadi}}{{Daughton} et~al.}{2006}]{daughton_06}
{Daughton} W.,  {Scudder} J.,   {Karimabadi} H.,  2006, \mn@doi [Physics of
  Plasmas] {10.1063/1.2218817}, \href
  {http://adsabs.harvard.edu/abs/2006PhPl...13g2101D} {13, 072101}

\bibitem[\protect\citeauthoryear{{Di Salvo}, {Done}, {{\.Z}ycki}, {Burderi}  \&
  {Robba}}{{Di Salvo} et~al.}{2001}]{DiSalvo+01}
{Di Salvo} T.,  {Done} C.,  {{\.Z}ycki} P.~T.,  {Burderi} L.,   {Robba} N.~R.,
  2001, \mn@doi [\apj] {10.1086/318396}, \href
  {https://ui.adsabs.harvard.edu/abs/2001ApJ...547.1024D} {547, 1024}

\bibitem[\protect\citeauthoryear{{Frontera} et~al.,}{{Frontera}
  et~al.}{2001}]{Frontera+01}
{Frontera} F.,  et~al., 2001, \mn@doi [\apj] {10.1086/318304}, \href
  {https://ui.adsabs.harvard.edu/abs/2001ApJ...546.1027F} {546, 1027}

\bibitem[\protect\citeauthoryear{{Galeev}, {Rosner}  \& {Vaiana}}{{Galeev}
  et~al.}{1979}]{Galeev+79}
{Galeev} A.~A.,  {Rosner} R.,   {Vaiana} G.~S.,  1979, \mn@doi [\apj]
  {10.1086/156957}, \href
  {https://ui.adsabs.harvard.edu/abs/1979ApJ...229..318G} {229, 318}

\bibitem[\protect\citeauthoryear{{Gary}}{{Gary}}{1992}]{Gary_92}
{Gary} S.~P.,  1992, \mn@doi [\jgr] {10.1029/92JA00299}, \href
  {https://ui.adsabs.harvard.edu/abs/1992JGR....97.8519G} {97, 8519}

\bibitem[\protect\citeauthoryear{{Gary}, {Li}, {O'Rourke}  \& {Winske}}{{Gary}
  et~al.}{1998}]{Gary+98}
{Gary} S.~P.,  {Li} H.,  {O'Rourke} S.,   {Winske} D.,  1998, \mn@doi [\jgr]
  {10.1029/98JA01174}, \href
  {https://ui.adsabs.harvard.edu/abs/1998JGR...10314567G} {103, 14567}

\bibitem[\protect\citeauthoryear{{Goodman} \& {Uzdensky}}{{Goodman} \&
  {Uzdensky}}{2008}]{Goodman&Uzdensky_08}
{Goodman} J.,  {Uzdensky} D.,  2008, \mn@doi [\apj] {10.1086/592345}, \href
  {https://ui.adsabs.harvard.edu/abs/2008ApJ...688..555G} {688, 555}

\bibitem[\protect\citeauthoryear{{Hakobyan}, {Petropoulou}, {Spitkovsky}  \&
  {Sironi}}{{Hakobyan} et~al.}{2020}]{hakobyan_20}
{Hakobyan} H.,  {Petropoulou} M.,  {Spitkovsky} A.,   {Sironi} L.,  2020, arXiv
  e-prints, \href {https://ui.adsabs.harvard.edu/abs/2020arXiv200612530H} {p.
  arXiv:2006.12530}

\bibitem[\protect\citeauthoryear{{Hall}}{{Hall}}{1979}]{Hall+79}
{Hall} A.~N.,  1979, \mn@doi [Journal of Plasma Physics]
  {10.1017/S0022377800022005}, \href
  {https://ui.adsabs.harvard.edu/abs/1979JPlPh..21..431H} {21, 431}

\bibitem[\protect\citeauthoryear{{Harris}}{{Harris}}{1962}]{1962NCim...23..115H}
{Harris} E.~G.,  1962, \mn@doi [Il Nuovo Cimento] {10.1007/BF02733547}, \href
  {https://ui.adsabs.harvard.edu/abs/1962NCim...23..115H} {23, 115}

\bibitem[\protect\citeauthoryear{{Hasegawa}}{{Hasegawa}}{1969}]{Hasegawa+69}
{Hasegawa} A.,  1969, \mn@doi [Physics of Fluids] {10.1063/1.1692407}, \href
  {https://ui.adsabs.harvard.edu/abs/1969PhFl...12.2642H} {12, 2642}

\bibitem[\protect\citeauthoryear{{Hellinger} \& {Matsumoto}}{{Hellinger} \&
  {Matsumoto}}{2000}]{Hellinger&Matsumoto_00}
{Hellinger} P.,  {Matsumoto} H.,  2000, \mn@doi [\jgr] {10.1029/1999JA000297},
  \href {https://ui.adsabs.harvard.edu/abs/2000JGR...10510519H} {105, 10519}

\bibitem[\protect\citeauthoryear{{Hellinger}, {Tr{\'a}vn{\'\i}{\v{c}}ek},
  {Kasper}  \& {Lazarus}}{{Hellinger} et~al.}{2006}]{Hellinger+06}
{Hellinger} P.,  {Tr{\'a}vn{\'\i}{\v{c}}ek} P.,  {Kasper} J.~C.,   {Lazarus}
  A.~J.,  2006, \mn@doi [\grl] {10.1029/2006GL025925}, \href
  {https://ui.adsabs.harvard.edu/abs/2006GeoRL..33.9101H} {33, L09101}

\bibitem[\protect\citeauthoryear{{Jiang}, {Stone}  \& {Davis}}{{Jiang}
  et~al.}{2019}]{jiang_19}
{Jiang} Y.-F.,  {Stone} J.~M.,   {Davis} S.~W.,  2019, \mn@doi [\apj]
  {10.3847/1538-4357/ab29ff}, \href
  {https://ui.adsabs.harvard.edu/abs/2019ApJ...880...67J} {880, 67}

\bibitem[\protect\citeauthoryear{{Kara} et~al.,}{{Kara} et~al.}{2019}]{Kara_19}
{Kara} E.,  et~al., 2019, \mn@doi [\nat] {10.1038/s41586-018-0803-x}, \href
  {https://ui.adsabs.harvard.edu/\#abs/2019Natur.565..198K} {565, 198}

\bibitem[\protect\citeauthoryear{{Kilian}, {Li}, {Guo}  \& {Li}}{{Kilian}
  et~al.}{2020}]{Kilian+20}
{Kilian} P.,  {Li} X.,  {Guo} F.,   {Li} H.,  2020, \mn@doi [\apj]
  {10.3847/1538-4357/aba1e9}, \href
  {https://ui.adsabs.harvard.edu/abs/2020ApJ...899..151K} {899, 151}

\bibitem[\protect\citeauthoryear{{Kunz}, {Schekochihin}  \& {Stone}}{{Kunz}
  et~al.}{2014}]{Kunz+14}
{Kunz} M.~W.,  {Schekochihin} A.~A.,   {Stone} J.~M.,  2014, \mn@doi [\prl]
  {10.1103/PhysRevLett.112.205003}, \href
  {https://ui.adsabs.harvard.edu/abs/2014PhRvL.112t5003K} {112, 205003}

\bibitem[\protect\citeauthoryear{{Loureiro}, {Schekochihin}  \&
  {Cowley}}{{Loureiro} et~al.}{2007}]{loureiro_07}
{Loureiro} N.~F.,  {Schekochihin} A.~A.,   {Cowley} S.~C.,  2007, \mn@doi
  [Physics of Plasmas] {10.1063/1.2783986}, \href
  {http://adsabs.harvard.edu/abs/2007PhPl...14j0703L} {14, 100703}

\bibitem[\protect\citeauthoryear{{McConnell} et~al.,}{{McConnell}
  et~al.}{2002}]{McConnell+02}
{McConnell} M.~L.,  et~al., 2002, \mn@doi [\apj] {10.1086/340436}, \href
  {https://ui.adsabs.harvard.edu/abs/2002ApJ...572..984M} {572, 984}

\bibitem[\protect\citeauthoryear{{Melzani}, {Winisdoerffer}, {Walder},
  {Folini}, {Favre}, {Krastanov}  \& {Messmer}}{{Melzani}
  et~al.}{2013}]{Melzani+13}
{Melzani} M.,  {Winisdoerffer} C.,  {Walder} R.,  {Folini} D.,  {Favre} J.~M.,
  {Krastanov} S.,   {Messmer} P.,  2013, \mn@doi [\aap]
  {10.1051/0004-6361/201321557}, \href
  {https://ui.adsabs.harvard.edu/abs/2013A&A...558A.133M} {558, A133}

\bibitem[\protect\citeauthoryear{{Melzani}, {Walder}, {Folini}, {Winisdoerffer}
   \& {Favre}}{{Melzani} et~al.}{2014a}]{Melzani+14a}
{Melzani} M.,  {Walder} R.,  {Folini} D.,  {Winisdoerffer} C.,   {Favre} J.~M.,
   2014a, \mn@doi [\aap] {10.1051/0004-6361/201424083}, \href
  {https://ui.adsabs.harvard.edu/abs/2014A&A...570A.111M} {570, A111}

\bibitem[\protect\citeauthoryear{{Melzani}, {Walder}, {Folini}, {Winisdoerffer}
   \& {Favre}}{{Melzani} et~al.}{2014b}]{Melzani+14b}
{Melzani} M.,  {Walder} R.,  {Folini} D.,  {Winisdoerffer} C.,   {Favre} J.~M.,
   2014b, \mn@doi [\aap] {10.1051/0004-6361/201424193}, \href
  {https://ui.adsabs.harvard.edu/abs/2014A&A...570A.112M} {570, A112}

\bibitem[\protect\citeauthoryear{{Melzani}, {Walder}, {Folini}, {Winisdoerffer}
   \& {Favre}}{{Melzani} et~al.}{2014c}]{melzani_14}
{Melzani} M.,  {Walder} R.,  {Folini} D.,  {Winisdoerffer} C.,   {Favre} J.~M.,
   2014c, \mn@doi [\aap] {10.1051/0004-6361/201424193}, \href
  {http://adsabs.harvard.edu/abs/2014A%26A...570A.112M} {570, A112}

\bibitem[\protect\citeauthoryear{{Nathanail}, {Mpisketzis}, {Porth}, {Fromm}
  \& {Rezzolla}}{{Nathanail} et~al.}{2021}]{Nathanail+21}
{Nathanail} A.,  {Mpisketzis} V.,  {Porth} O.,  {Fromm} C.~M.,   {Rezzolla} L.,
   2021, arXiv e-prints, \href
  {https://ui.adsabs.harvard.edu/abs/2021arXiv211103689N} {p. arXiv:2111.03689}

\bibitem[\protect\citeauthoryear{{Ni}, {Ziegler}, {Huang}, {Lin}  \&
  {Mei}}{{Ni} et~al.}{2012}]{Ni+12}
{Ni} L.,  {Ziegler} U.,  {Huang} Y.-M.,  {Lin} J.,   {Mei} Z.,  2012, \mn@doi
  [Physics of Plasmas] {10.1063/1.4736993}, \href
  {https://ui.adsabs.harvard.edu/abs/2012PhPl...19g2902N} {19, 072902}

\bibitem[\protect\citeauthoryear{{Petropoulou} \& {Sironi}}{{Petropoulou} \&
  {Sironi}}{2018}]{petropoulou_18}
{Petropoulou} M.,  {Sironi} L.,  2018, \mn@doi [\mnras]
  {10.1093/mnras/sty2702}, \href
  {http://adsabs.harvard.edu/abs/2018MNRAS.481.5687P} {481, 5687}

\bibitem[\protect\citeauthoryear{{Petropoulou}, {Sironi}, {Spitkovsky}  \&
  {Giannios}}{{Petropoulou} et~al.}{2019}]{petropoulou_19}
{Petropoulou} M.,  {Sironi} L.,  {Spitkovsky} A.,   {Giannios} D.,  2019,
  \mn@doi [\apj] {10.3847/1538-4357/ab287a}, \href
  {https://ui.adsabs.harvard.edu/abs/2019ApJ...880...37P} {880, 37}

\bibitem[\protect\citeauthoryear{{Quest} \& {Shapiro}}{{Quest} \&
  {Shapiro}}{1996}]{Quest&Shapiro_96}
{Quest} K.~B.,  {Shapiro} V.~D.,  1996, \mn@doi [\jgr] {10.1029/96JA01534},
  \href {https://ui.adsabs.harvard.edu/abs/1996JGR...10124457Q} {101, 24457}

\bibitem[\protect\citeauthoryear{{Ripperda}, {Bacchini}  \&
  {Philippov}}{{Ripperda} et~al.}{2020}]{ripperda_20}
{Ripperda} B.,  {Bacchini} F.,   {Philippov} A.~A.,  2020, \mn@doi [\apj]
  {10.3847/1538-4357/ababab}, \href
  {https://ui.adsabs.harvard.edu/abs/2020ApJ...900..100R} {900, 100}

\bibitem[\protect\citeauthoryear{{Ripperda}, {Liska}, {Chatterjee}, {Musoke},
  {Philippov}, {Markoff}, {Tchekhovskoy}  \& {Younsi}}{{Ripperda}
  et~al.}{2021}]{ripperda_21}
{Ripperda} B.,  {Liska} M.,  {Chatterjee} K.,  {Musoke} G.,  {Philippov} A.~A.,
   {Markoff} S.~B.,  {Tchekhovskoy} A.,   {Younsi} Z.,  2021, arXiv e-prints,
  \href {https://ui.adsabs.harvard.edu/abs/2021arXiv210915115R} {p.
  arXiv:2109.15115}

\bibitem[\protect\citeauthoryear{{Riquelme}, {Quataert}  \&
  {Verscharen}}{{Riquelme} et~al.}{2015}]{Riquelme+15}
{Riquelme} M.~A.,  {Quataert} E.,   {Verscharen} D.,  2015, \mn@doi [\apj]
  {10.1088/0004-637X/800/1/27}, \href
  {https://ui.adsabs.harvard.edu/abs/2015ApJ...800...27R} {800, 27}

\bibitem[\protect\citeauthoryear{{Rowan}, {Sironi}  \& {Narayan}}{{Rowan}
  et~al.}{2017}]{rowan_17}
{Rowan} M.~E.,  {Sironi} L.,   {Narayan} R.,  2017, \mn@doi [\apj]
  {10.3847/1538-4357/aa9380}, \href
  {http://adsabs.harvard.edu/abs/2017ApJ...850...29R} {850, 29}

\bibitem[\protect\citeauthoryear{{Rowan}, {Sironi}  \& {Narayan}}{{Rowan}
  et~al.}{2019}]{rowan_19}
{Rowan} M.~E.,  {Sironi} L.,   {Narayan} R.,  2019, \mn@doi [\apj]
  {10.3847/1538-4357/ab03d7}, \href
  {http://adsabs.harvard.edu/abs/2019ApJ...873....2R} {873, 2}

\bibitem[\protect\citeauthoryear{{Samsonov}, {Pudovkin}, {Gary}  \&
  {Hubert}}{{Samsonov} et~al.}{2001}]{Samsonov+01}
{Samsonov} A.~A.,  {Pudovkin} M.~I.,  {Gary} S.~P.,   {Hubert} D.,  2001,
  \mn@doi [\jgr] {10.1029/2000JA900150}, \href
  {https://ui.adsabs.harvard.edu/abs/2001JGR...10621689S} {106, 21689}

\bibitem[\protect\citeauthoryear{{Sharma}, {Hammett}, {Quataert}  \&
  {Stone}}{{Sharma} et~al.}{2006}]{Sharma+06}
{Sharma} P.,  {Hammett} G.~W.,  {Quataert} E.,   {Stone} J.~M.,  2006, \mn@doi
  [\apj] {10.1086/498405}, \href
  {https://ui.adsabs.harvard.edu/abs/2006ApJ...637..952S} {637, 952}

\bibitem[\protect\citeauthoryear{{Sironi}}{{Sironi}}{2022}]{sironi_22}
{Sironi} L.,  2022, \mn@doi [\prl] {10.1103/PhysRevLett.128.145102}, \href
  {https://ui.adsabs.harvard.edu/abs/2022PhRvL.128n5102S} {128, 145102}

\bibitem[\protect\citeauthoryear{{Sironi} \& {Beloborodov}}{{Sironi} \&
  {Beloborodov}}{2020}]{Sironi_20}
{Sironi} L.,  {Beloborodov} A.~M.,  2020, \mn@doi [\apj]
  {10.3847/1538-4357/aba622}, \href
  {https://ui.adsabs.harvard.edu/abs/2020ApJ...899...52S} {899, 52}

\bibitem[\protect\citeauthoryear{{Sironi} \& {Narayan}}{{Sironi} \&
  {Narayan}}{2015}]{Sironi&Narayan_15}
{Sironi} L.,  {Narayan} R.,  2015, \mn@doi [\apj] {10.1088/0004-637X/800/2/88},
  \href {https://ui.adsabs.harvard.edu/abs/2015ApJ...800...88S} {800, 88}

\bibitem[\protect\citeauthoryear{{Sironi}, {Petropoulou}  \&
  {Giannios}}{{Sironi} et~al.}{2015}]{sironi_15}
{Sironi} L.,  {Petropoulou} M.,   {Giannios} D.,  2015, \mn@doi [\mnras]
  {10.1093/mnras/stv641}, \href
  {http://adsabs.harvard.edu/abs/2015MNRAS.450..183S} {450, 183}

\bibitem[\protect\citeauthoryear{{Sironi}, {Giannios}  \&
  {Petropoulou}}{{Sironi} et~al.}{2016}]{sironi_16}
{Sironi} L.,  {Giannios} D.,   {Petropoulou} M.,  2016, \mn@doi [\mnras]
  {10.1093/mnras/stw1620}, \href
  {http://adsabs.harvard.edu/abs/2016MNRAS.462...48S} {462, 48}

\bibitem[\protect\citeauthoryear{{Spitkovsky}}{{Spitkovsky}}{2005}]{spitkovsky_05}
{Spitkovsky} A.,  2005, in {T.~Bulik, B.~Rudak, \& G.~Madejski} ed.,  AIP Conf.
  Ser. Vol. 801, Astrophysical Sources of High Energy Particles and Radiation.
  p.~345 (\mn@eprint {} {arXiv:astro-ph/0603211}), \mn@doi{10.1063/1.2141897}

\bibitem[\protect\citeauthoryear{{Spitzer}}{{Spitzer}}{1965}]{Spitzer_65}
{Spitzer} L.,  1965, {Physics of fully ionized gases}

\bibitem[\protect\citeauthoryear{{Sridhar}, {Garc{\'\i}a}, {Steiner},
  {Connors}, {Grinberg}  \& {Harrison}}{{Sridhar} et~al.}{2020}]{Sridhar+20}
{Sridhar} N.,  {Garc{\'\i}a} J.~A.,  {Steiner} J.~F.,  {Connors} R. M.~T.,
  {Grinberg} V.,   {Harrison} F.~A.,  2020, \mn@doi [\apj]
  {10.3847/1538-4357/ab64f5}, \href
  {https://ui.adsabs.harvard.edu/abs/2020ApJ...890...53S} {890, 53}

\bibitem[\protect\citeauthoryear{{Sridhar}, {Sironi}  \&
  {Beloborodov}}{{Sridhar} et~al.}{2021}]{Sridhar+21c}
{Sridhar} N.,  {Sironi} L.,   {Beloborodov} A.~M.,  2021, \mn@doi [\mnras]
  {10.1093/mnras/stab2534}, \href
  {https://ui.adsabs.harvard.edu/abs/2021MNRAS.507.5625S} {507, 5625}

\bibitem[\protect\citeauthoryear{{Stepney}}{{Stepney}}{1983}]{Stepney_83}
{Stepney} S.,  1983, \mn@doi [\mnras] {10.1093/mnras/202.2.467}, \href
  {https://ui.adsabs.harvard.edu/abs/1983MNRAS.202..467S} {202, 467}

\bibitem[\protect\citeauthoryear{{Tajima} \& {Shibata}}{{Tajima} \&
  {Shibata}}{1997}]{tajima_shibata_97}
{Tajima} T.,  {Shibata} K.,  eds, 1997, {Plasma astrophysics}

\bibitem[\protect\citeauthoryear{{Tamburini}, {Pegoraro}, {Di Piazza}, {Keitel}
   \& {Macchi}}{{Tamburini} et~al.}{2010}]{2010NJPh...12l3005T}
{Tamburini} M.,  {Pegoraro} F.,  {Di Piazza} A.,  {Keitel} C.~H.,   {Macchi}
  A.,  2010, \mn@doi [New Journal of Physics] {10.1088/1367-2630/12/12/123005},
  \href {https://ui.adsabs.harvard.edu/abs/2010NJPh...12l3005T} {12, 123005}

\bibitem[\protect\citeauthoryear{{Uzdensky}, {Loureiro}  \&
  {Schekochihin}}{{Uzdensky} et~al.}{2010}]{uzdensky_10}
{Uzdensky} D.~A.,  {Loureiro} N.~F.,   {Schekochihin} A.~A.,  2010, \mn@doi
  [Physical Review Letters] {10.1103/PhysRevLett.105.235002}, \href
  {http://adsabs.harvard.edu/abs/2010PhRvL.105w5002U} {105, 235002}

\bibitem[\protect\citeauthoryear{{Vay}}{{Vay}}{2008}]{Vay_08}
{Vay} J.-L.,  2008, \mn@doi [Physics of Plasmas] {10.1063/1.2837054}, \href
  {https://ui.adsabs.harvard.edu/abs/2008PhPl...15e6701V} {15, 056701}

\bibitem[\protect\citeauthoryear{{Wang} et~al.,}{{Wang} et~al.}{2021}]{Wang+21}
{Wang} J.,  et~al., 2021, \mn@doi [\apjl] {10.3847/2041-8213/abec79}, \href
  {https://ui.adsabs.harvard.edu/abs/2021ApJ...910L...3W} {910, L3}

\bibitem[\protect\citeauthoryear{{Werner}, {Uzdensky}, {Begelman}, {Cerutti}
  \& {Nalewajko}}{{Werner} et~al.}{2018}]{werner_18}
{Werner} G.~R.,  {Uzdensky} D.~A.,  {Begelman} M.~C.,  {Cerutti} B.,
  {Nalewajko} K.,  2018, \mn@doi [\mnras] {10.1093/mnras/stx2530}, \href
  {http://adsabs.harvard.edu/abs/2018MNRAS.473.4840W} {473, 4840}

\bibitem[\protect\citeauthoryear{{Werner}, {Philippov}  \& {Uzdensky}}{{Werner}
  et~al.}{2019}]{werner_19}
{Werner} G.~R.,  {Philippov} A.~A.,   {Uzdensky} D.~A.,  2019, \mn@doi [\mnras]
  {10.1093/mnrasl/sly157}, \href
  {http://adsabs.harvard.edu/abs/2019MNRAS.482L..60W} {482, L60}

\bibitem[\protect\citeauthoryear{{Zdziarski} \& {Gierli{\'n}ski}}{{Zdziarski}
  \& {Gierli{\'n}ski}}{2004}]{Zdziarski&Gierlinski_04}
{Zdziarski} A.~A.,  {Gierli{\'n}ski} M.,  2004, \mn@doi [Progress of
  Theoretical Physics Supplement] {10.1143/PTPS.155.99}, \href
  {https://ui.adsabs.harvard.edu/abs/2004PThPS.155...99Z} {155, 99}

\bibitem[\protect\citeauthoryear{{Zdziarski}, {Malyshev}, {Chernyakova}  \&
  {Pooley}}{{Zdziarski} et~al.}{2017}]{Zdziarski+17}
{Zdziarski} A.~A.,  {Malyshev} D.,  {Chernyakova} M.,   {Pooley} G.~G.,  2017,
  \mn@doi [\mnras] {10.1093/mnras/stx1846}, \href
  {https://ui.adsabs.harvard.edu/abs/2017MNRAS.471.3657Z} {471, 3657}

\bibitem[\protect\citeauthoryear{{Zhang}, {Sironi}  \& {Giannios}}{{Zhang}
  et~al.}{2021}]{zhang_21}
{Zhang} H.,  {Sironi} L.,   {Giannios} D.,  2021, \mn@doi [\apj]
  {10.3847/1538-4357/ac2e08}, \href
  {https://ui.adsabs.harvard.edu/abs/2021ApJ...922..261Z} {922, 261}

\bibitem[\protect\citeauthoryear{{Zhdankin}, {Uzdensky}  \& {Kunz}}{{Zhdankin}
  et~al.}{2020}]{zhdankin_20b}
{Zhdankin} V.,  {Uzdensky} D.~A.,   {Kunz} M.~W.,  2020, arXiv e-prints, \href
  {https://ui.adsabs.harvard.edu/abs/2020arXiv200712050Z} {p. arXiv:2007.12050}

\makeatother
\end{thebibliography}

%%%%%%%%%%%%%%%%%%%%%%%%%%%%%%%%%%%%%%%%%%%%%%%%%%

%%%%%%%%%%%%%%%%% APPENDICES %%%%%%%%%%%%%%%%%%%%%

\appendix

\section{Simulation parameters}\label{appendix:table}

We present in Table \ref{tab:setup} the input parameters of all the simulations presented in the main text. 

\begin{table}
\setlength{\tabcolsep}{2.5pt}
\begin{center}
      \caption{List of numerical and physical input parameters.}
      \label{tab:setup}
      \begin{tabular}{ccccc}
       \hline
       \hline
      $\sigmai^{[1]}$ & $\gamma_{\rm cr}^{[2]}$ & Domain size$^{[3]}$ & $\gcool^{[4]}$ & $\mi/\me^{[5]}$ \\ 
      & & $\Lx/(\compe)$ & & \\ % <--
       \hline
      3 & $\infty$  & 1680 & $\infty$ & 7  \\ % <--
      3 & $\infty$  & 1680 & $\infty$ & 29  \\ % <--
      3 & $\infty$  & 1680  & $\infty$ & 115  \\ % <--
      3 & $\infty$  & 1680  & $\infty$ & 459  \\ % <--
      $^\dagger$3 & $\infty$  & 1680  & $\infty$ & 1836 \\ % <--
      $^\ddagger$3 & 13.7  & 840  & 0.24 & 29 \\ % <--
      3 & 19.4  & 1680  & 0.24 & 29 \\ % <--
      3 & 22.5  & 1680  & 0.33 & 29 \\ % <--
      3 & 32  & 1680  & 0.66 & 29 \\ % <--
      3 & 45  & 1680  & 1.32 & 29 \\ % <--
      3 & 22.5  & 3360  & 0.16 & 29 \\ % <--
      3 & 19.4  & 3360  & 0.12 & 29 \\ % <--
      3 & 19.4  & 6720  & 0.06 & 29 \\ % <--
       \hline
      2 & 17.5  & 1680  & 0.24 & 29 \\ % <--
      2 & 17.5  & 3360  & 0.12 & 29 \\ % <--
      2 & 17.5  & 6720  & 0.06 & 29 \\ % <--
       \hline
      1 & $\infty$  & 1680  & $\infty$ & 7   \\ % <--
      1 & $\infty$  & 1680  & $\infty$ & 29  \\ % <--
      1 & $\infty$  & 1680 & $\infty$ & 115  \\ % <--
      1 & $\infty$  & 1680  & $\infty$ & 459  \\ % <--
      1 & $\infty$  & 1680  & $\infty$ & 1836 \\ % <--
      1 & 12.1  & 420  & 0.66 & 29 \\ % <--
      1 & 17.2  & 420  & 1.32 & 29 \\ % <--
      1 & 10.4  & 840  & 0.24 & 29 \\ % <--
      1 & 10.4  & 1680  & 0.12 & 29 \\ % <--
      1 & 10.4  & 3360  & 0.06 & 29 \\ % <--

	 \hline
    \end{tabular}
      \begin{tablenotes}
      \footnotesize
      \item \textit{Note.} {All simulations are performed for the same duration of $t_{\rm sim}\sim4.8\Lx/v_{\rm A}$, with the same initial particle number density, $n_0=4$, and spatial resolution ${\cal R}=(\compe)/\delta=5$, unless specified otherwise. The description of each column is as follows. $^{[1]}$Magnetization in the upstream plasma; $^{[2]}$Critical Lorentz factor---a proxy for the intensity of incident photon field ($\gamma_{\rm cr}=\infty$ implies no IC cooling, and smaller $\gamma_{\rm cr}$ implies greater IC cooling; see Eq.~\ref{sec:IC_cooling});  $^{[3]}$Half-length of the computational domain $\Lx$, along the $x$-direction, in units of electron plasma skin depth; $^{[4]}$ $\gcool$, as defined in Eq.~\ref{eqn:varrho}; $^{[5]}$Ion to electron mass ratio. $^\dagger$ indicates the model with three different available values of ${\cal R}=2.5,5$ and 10 (see Appendix~\ref{appendix:resolution_convergence}). $^\ddagger$ indicates the model with two different available values of $n_0=4$ and 64, different number of passes of current-density filtering (8 and 32) (see Appendix~\ref{appendix:ppc_ntimes_convergence}), and different initial temperatures $\theta_{\rm e}=0.02$ and $10^{-4}$ (see Appendix~\ref{appendix:resolution_convergence}).}
      \end{tablenotes}
\end{center}
\end{table}

\section{Dependence on the mass ratio} \label{appendix:mass_ratio}

\begin{figure}
\includegraphics[width=8cm]{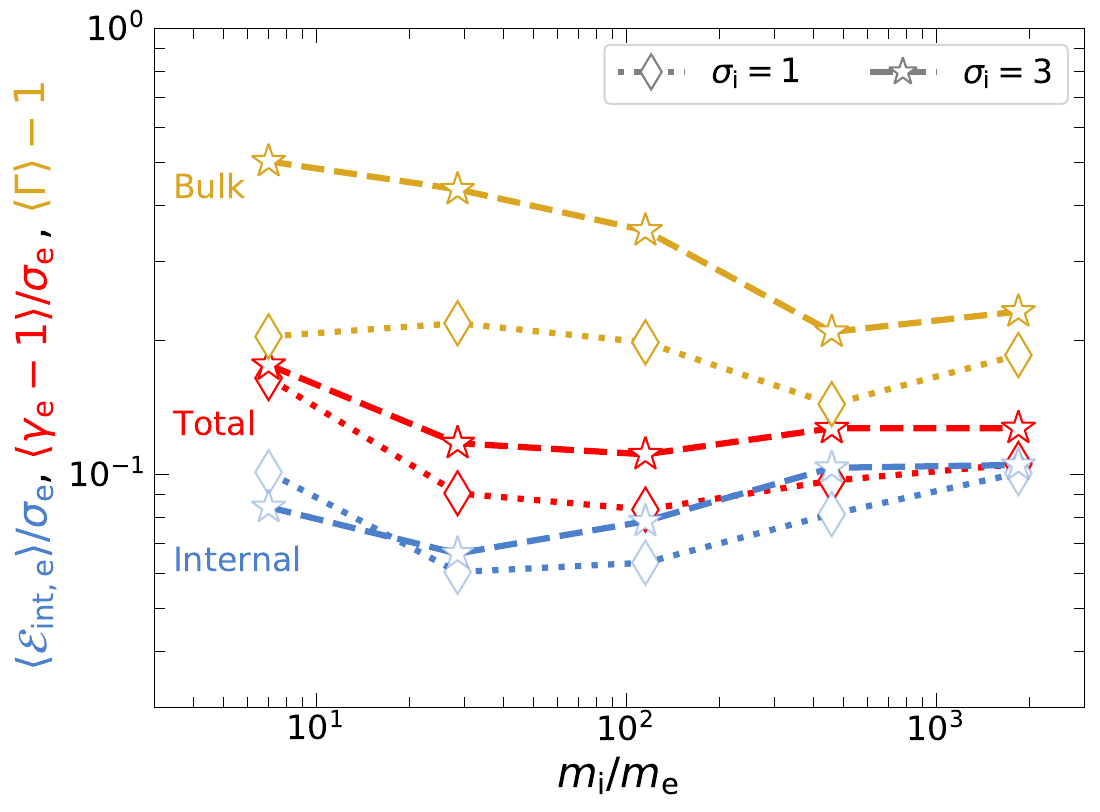}
\caption{Mean energies of electrons (in units of $\me c^2$) for different $\mi/\me$ and $\sigmai$. The red markers indicate the total energy $\langle\gammae-1\rangle$, the blue markers indicate the internal energy $\langle\varepsilon_{\rm int,e}\rangle$, and the golden-brown markers indicate the bulk motion's contribution $\langle\Gamma\rangle-1$. The total and internal energies are normalized by $\sigmae$. The models with $\sigmai=1$ and $\sigmai=3$ are denoted by star and diamond markers, connected by dashed and dotted lines, respectively.}
\label{fig:energies_mime_sigma}
\end{figure}

\begin{figure}
\includegraphics[width=8cm]{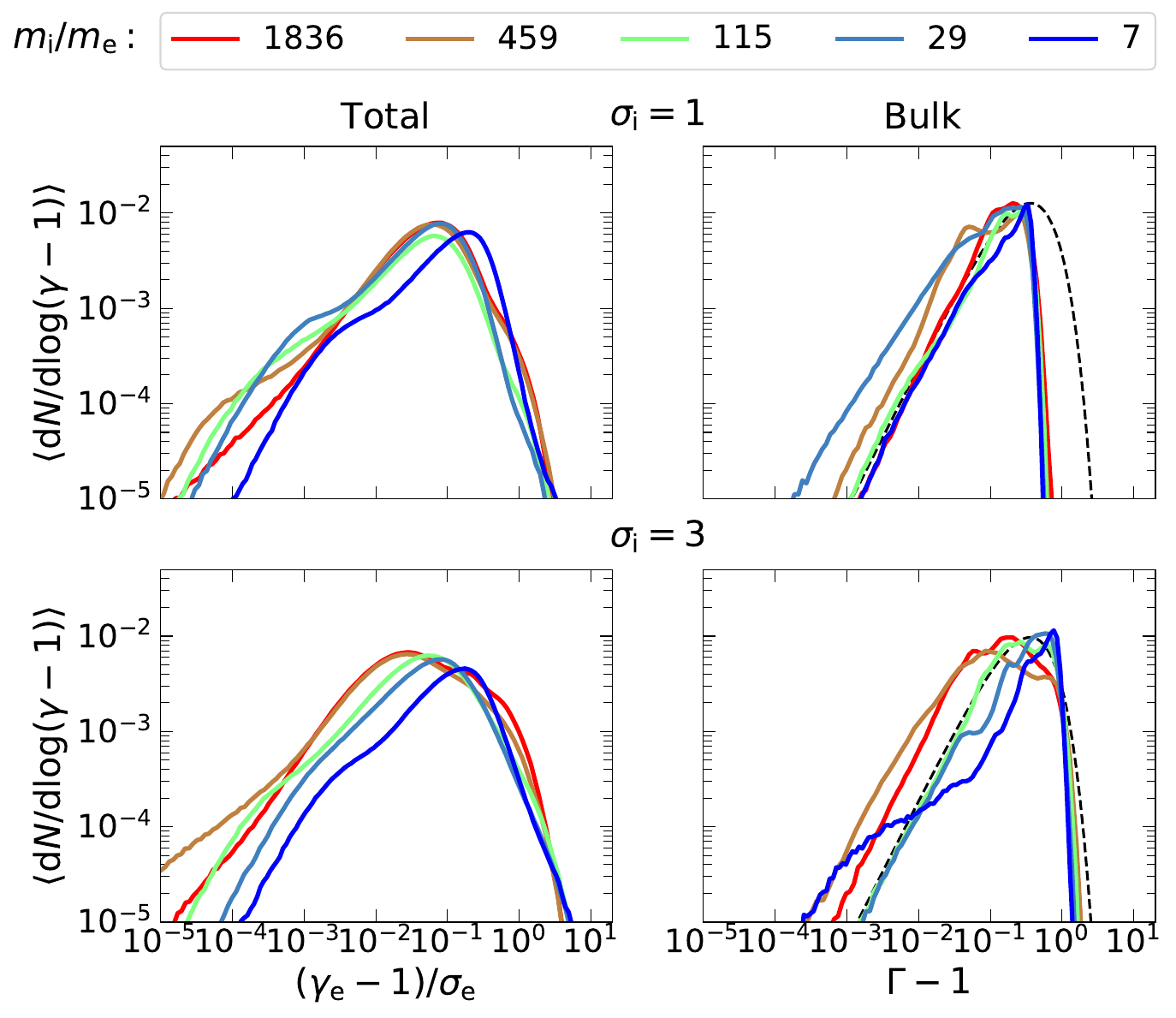}
\caption{Total (left column) and bulk (right column) energy spectra of electrons for models with no external IC cooling ($\gcool=\infty$), with different magnetizations $\sigmai=1$ (top row) and $\sigmai=3$ (bottom row), and different mass ratios $\mi/\me$ (color coded). For comparison, we also plot a Maxwellian distribution with  a temperature of 100\,keV (dashed black; right panels). The spectra are time-averaged in the interval $2 \lesssim T/(\Lx/v_{\rm A}) \lesssim 5$, where $\Lx/(\compe)=1680$ is the simulation domain size.}
\label{fig:spectra_mime_sigma}
\end{figure}

In this section, we check the influence of the mass ratio $\mi/\me$ on particle energization, for different $\sigmai$, in the absence of IC cooling. As we have discussed in the main text, the overall dynamics of the reconnection layer, including the rate of reconnection, is nearly independent of $\mi/\me$ (see \S\ref{subsec:general} and Fig.~\ref{fig:recrate}). In Fig.~\ref{fig:energies_mime_sigma}, we show that the mean total energy of electrons $\langle\gammae-1\rangle$ stays constant for a wide range of mass ratios, $7\lesssim\mi/\me\lesssim1836$. A modest increase by a factor of ${\sim}3$ is seen in the mean bulk energy $\langle\Gamma\rangle-1$ while $\mi/\me$ decreases by a factor of ${\sim}260$. Furthermore, this is observable only for $\sigmai\sim3$, and not for $\sigmai=1$ (Fig.~\ref{fig:spectra_mime_sigma}). For the purposes of our study, the choice of an artificially low mass ratio $\mi/\me=29$, as compared to its realistic value $\mi/\me=1836$, is not expected to impair the prospects of cold-chain Comptonization due to bulk motions. However, it is still valuable to check whether modest quantitative changes in the shape of the X-ray spectrum may come with higher, realistic mass ratios. This is left for future work.

\section{Convergence with respect to spatial resolution}\label{appendix:resolution_convergence}

\begin{figure}
\includegraphics[width=8cm]{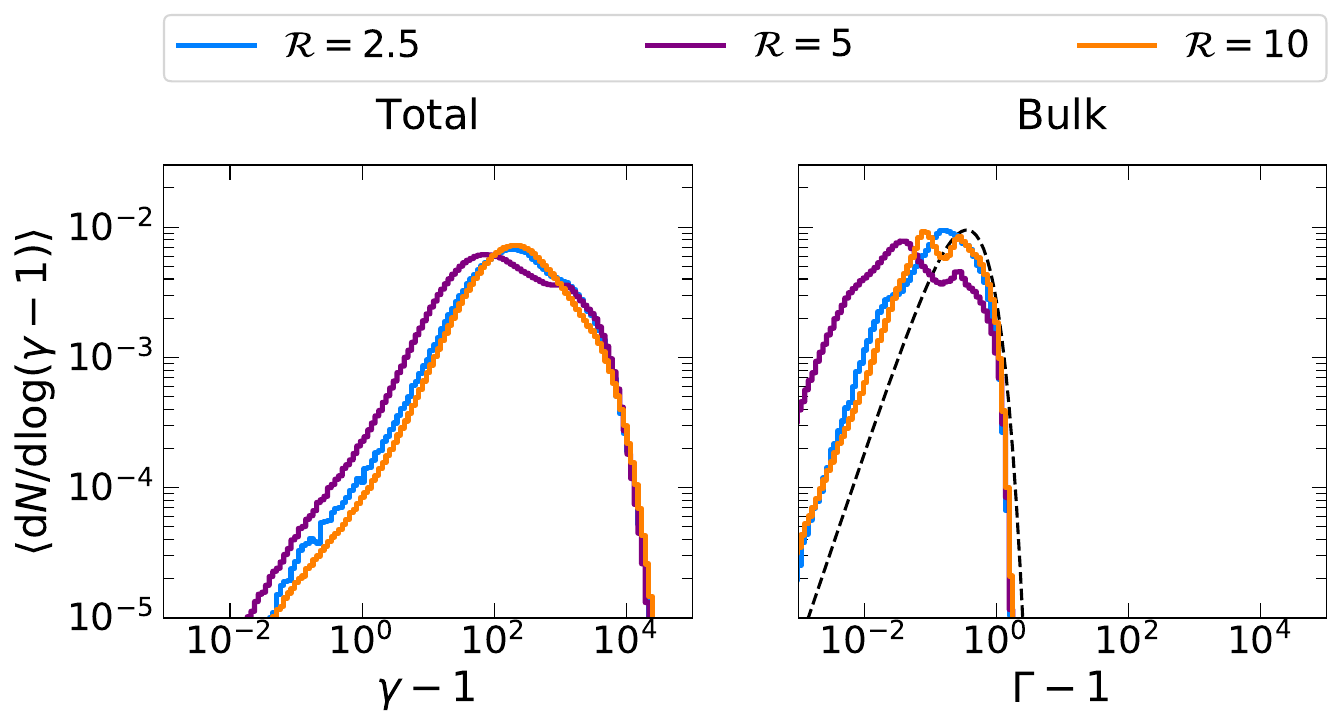}
\caption{Total (left panel) and bulk (right panel) energy spectra of electrons for models with the same $\sigmai=3$, mass ratio $\mi/\me=1836$, no external IC cooling ($\gcool=\infty$), with different resolutions ${\cal R}=2.5$ (blue), ${\cal R}=5$ (maroon; default value in this work) and ${\cal R}=10$ (orange). For comparison, we also plot a Maxwellian distribution with a temperature of 100\,keV (dashed black; right panels). The spectra are time-averaged in the interval $2 \lesssim T/(\Lx/v_{\rm A}) \lesssim 5$, where $\Lx/(\compe)=1680$ is the simulation domain size.}
\label{fig:spectra_resolution_1}
\end{figure}

Here, we check if the spatial resolution of the simulation domain influences the electrons' spectrum. We demonstrate with Fig.~\ref{fig:spectra_resolution_1} that the shape of the electrons' total energy spectrum and its bulk component do not change in a systematic manner with a change in spatial resolution in the range $2.5\le{\cal R}\le10$. In fact, the spectra for ${\cal R}=2.5$ and ${\cal R}=10$ appear to overlap remarkably well. We note that the results from our simulations---performed with the default value of ${\cal R}=5$ with outflow boundary conditions along $x$-direction---are robust with respect to the spatial resolution.

On the other hand, the Debye length $\lambda_{\rm De}=\sqrt{k_{\rm B}T_{\rm e}/4\pi n_0 e^2}$ under the initial plasma conditions of our fiducial simulations is only 0.05 cells. We assess whether our results are prone to any numerical artifacts due to under-resolved $\lambda_{\rm De}$. Overall, our aim is to perform simulations in a cold plasma, meaning where the initial thermal energy is much smaller than the rest mass energy. In this case, we argue that the reconnection physics does not depend on the specific choice of initial temperature. We perform new  simulations of IC-cooled reconnection, where $\lambda_{\rm De}$ is resolved. Rather than resolving $\lambda_{\rm De}$ by increasing ${\cal R}$---which is computationally very demanding (for our requirement of low $\gcool$, and so, a large $L_{\rm x}/(c/\omega_{\rm pe})$)---we initialize a relatively hotter upstream plasma: instead of our fiducial upstream temperature of $\theta_{\rm e}=k_{\rm B} T_{\rm e}/m_{\rm e}c^2 = 10^{-4}$, we now consider a temperature of $\theta_{\rm e} = 0.02$. While this is still cold enough compared to the energies particles attain as a result of reconnection (which is the regime we want to be in, as prefaced above), this choice of temperature also resolves $\lambda_{\rm De}$ with 1 cell assuming the same fiducial resolution of ${\cal R} = 5$. We have performed four new simulations with periodic boundary conditions along the $x$-direction (the boundary conditions along the $y$-direction remain the same as in our fiducial case) with the following parameters: ${\cal R} = 5, \sigmai = 3, n_0 = 4$, 32 rounds of current filtering, and $\mi/\me=29$. The parameters that differentiate the four models are: (1) $\gcool=0.24, \theta_{\rm e}=10^{-4}$ (as in the main text), (2) $\gcool=0.24, \theta_{\rm e}=0.02$, (3) $\gcool=\infty, \theta_{\rm e}=10^{-4}$ (no cooling), and (4) $\gcool=\infty, \theta_{\rm e}=0.02$ (no cooling). Note that in Appendix~\ref{appendix:recrate}, we show that these choices of simulation parameters yield a similar $\eta_{\rm rec}$ to our fiducial simulations.

\begin{figure}
\includegraphics[width=8.2cm]{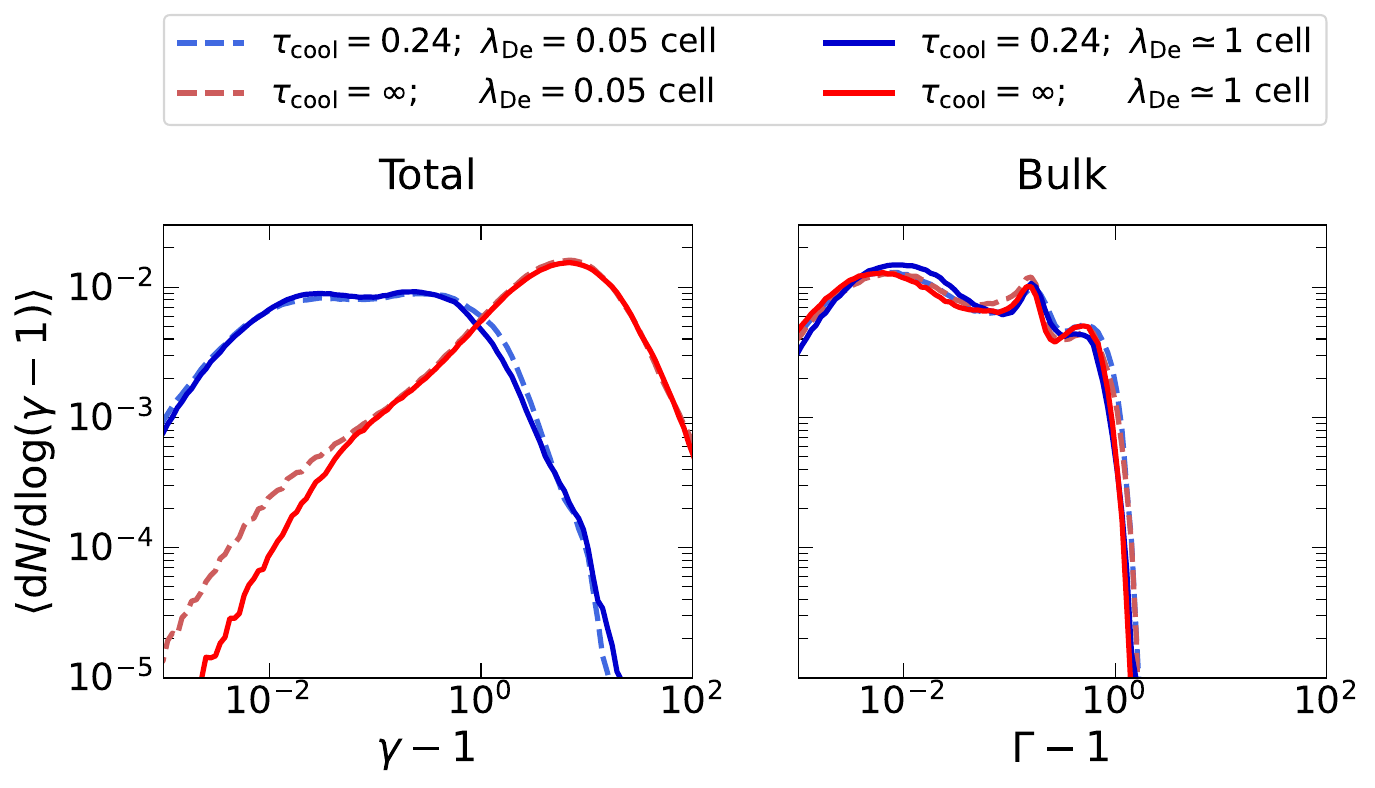}
\caption{Total (left panel) and bulk (right panel) energy spectra of electrons for models with the same $\sigmai=3$, $\mi/\me=29$, spatial resolution ${\cal R}=5$, but with different strengths of IC cooling, $\gcool=0.24$ (blue curves) and $\gcool=\infty$ (red curves), and Debye lengths $\lambda_{\rm De}=\sqrt{k_{\rm B}T_{\rm e}/4\pi n_0 e^2}$ = 0.05 cells (dashed curves) and 1 cell (solid curves). The spectra are time-averaged in the interval $2 \lesssim T/(\Lx/v_{\rm A}) \lesssim 5$, where $\Lx/(\compe)=840$ is the simulation domain size.}
\label{fig:spectra_resolution_2}
\end{figure}

Our results are presented in  Fig.~\ref{fig:spectra_resolution_2}, which shows the total energy spectrum (left panel) and the bulk component (right panel). One can see a close match in the spectral shape, location of the peak, its normalization, and the cut-off energy, for our current simulations with under-resolved $\lambda_{\rm De}$ (dashed curves) vs. the cases that resolve $\lambda_{\rm De}$ with 1 cell (solid curves). This similarity holds for both cooled (blue curves) and uncooled cases (red curves). The presence of a broad peak at $\Gamma-1 \sim 4\times10^{-3}$ in the bulk spectra is due to the formation of near-static monster plasmoids at the edges of the box due to the assumed periodic boundary conditions (here, plasmoids are not advected out of the box, as in our fiducial setup, but instead they merge and keep growing over time).

\section{Attainment of quasi-steady state} \label{appendix:recrate}

\begin{figure*}
\includegraphics[width=18cm]{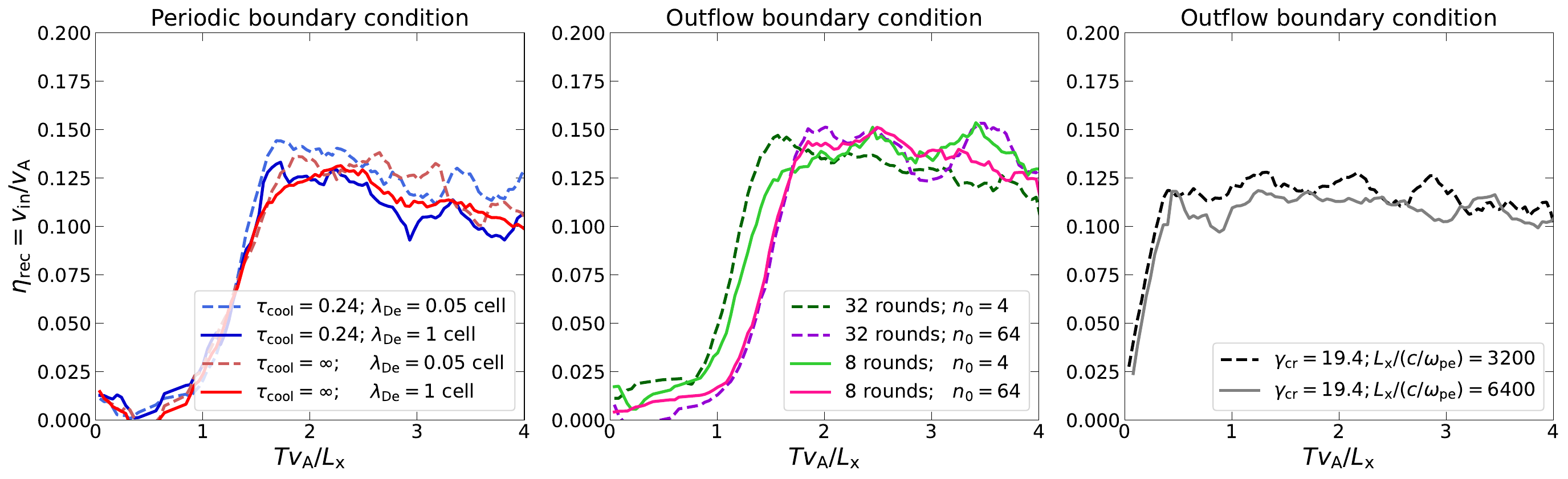}
\caption{Reconnection rate ($\eta_{\rm rec}$) under different simulation conditions. The fiducial set of simulation parameters are: ${\cal R} = 5, \sigmai = 3, n_0 = 4, \mi/\me=29, L_{\rm x}/(c/\omega_{\rm pe}) = 840, \gcool=24, \theta_{\rm e}=10^{-4} (\lambda_{\rm De}=\sqrt{k_{\rm B}T_{\rm e}/4\pi n_0 e^2}=0.05\,{\rm cells})$, 32 rounds of binomial current density filtering, and outflow boundary conditions along the $x$-direction. Each curve denotes a simulation setup that differs from the fiducial case by one or two parameters (bottom-right legend in each panel). In addition, we change the initial thickness of the current sheet $\Delta_{\rm y}$ to delay the onset of reconnection by $\sim1\,L{\rm x}/v_{\rm A}$, for simulations in the left and middle panels. Left panel: simulations that employ periodic boundary conditions along the $x$-direction, for different initial temperatures (equivalent to different $\lambda_{\rm De}$), in the presence and absence of IC cooling. Middle panel: simulations with different upstream particle number densities and rounds of binomial current filtering. Right panel: simulations with different strengths of IC cooling, $\gcool=0.06$ (grey solid curve) and $\gcool=0.12$ (black dashed curve).}
\label{fig:recrate_time}
\end{figure*}

All the quantities of interest to this paper are calculated for times $T/(\Lx/v_{\rm A}) \gtrsim 2$, for this is when the reconnection layer attains a quasi-steady state. i.e., the formation of plasmoids and their advection out of the domain occur at a somewhat steady rate. This happens concurrently with the saturation of the reconnection rate ($\eta_{\rm rec}=v_{\rm in}/v_{\rm A}$, which measures the inflow speed of upstream plasma into the reconnection layer). This is shown in Fig.~\ref{fig:recrate_time} for various simulation conditions. We also change the initial thickness of the current sheet to delay the onset of reconnection to see if this changes the way the system evolves once reconnection is triggered. Overall, we see that the layer attains a quasi-steady state $\sim1\,L_{\rm x}/v_{\rm A}$ after  reconnection is triggered, regardless of our chosen initial conditions.

\section{Convergence with respect to particle density and rounds of current density filtering} \label{appendix:ppc_ntimes_convergence}

\begin{figure}
\includegraphics[width=8cm]{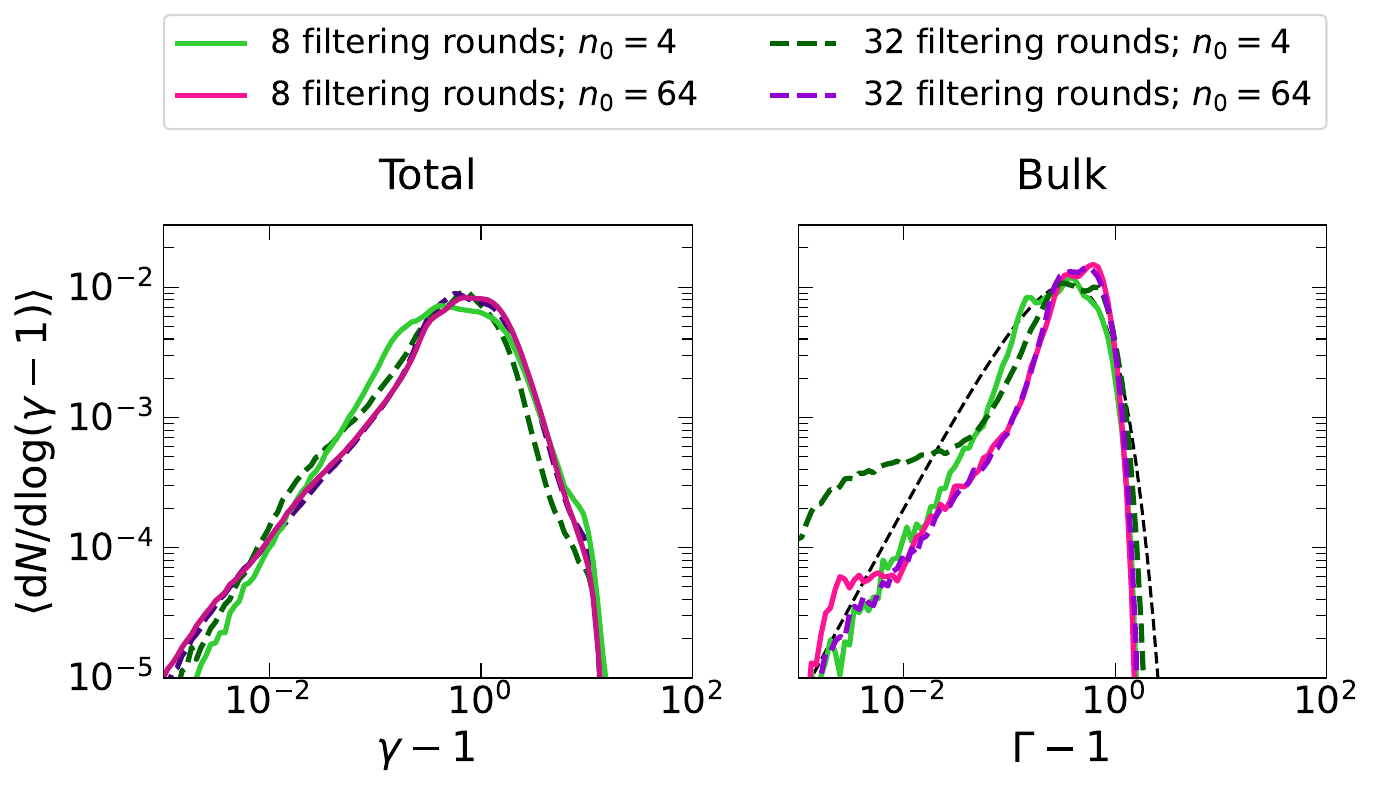}
\caption{Total (left panel) and bulk (right panel) energy spectra of electrons for models with the same $\sigmai=3$, $\mi/\me=29$, $\gcool=0.24$, but with different choices of particle number density $n_0=4$ (green curves; default value in this work) and $n_0=64$ (purple). We also show spectra from simulations with different number of binomial current density filtering: 32 rounds (dashed curves; default value in this work), and 8 rounds (solid curves). For comparison, we also plot a Maxwellian distribution with a temperature of 100\,keV (dashed black; right panel). The spectra are time-averaged in the interval $2 \lesssim T/(\Lx/v_{\rm A}) \lesssim 5$, where $\Lx/(\compe)=840$ is the simulation domain size.}
\label{fig:spectra_ppc0_ntimes}
\end{figure}

Throughout our simulations, we initialize the domain with four particles per cell ($n_0=4$; two electrons and two ions). Here, we check if our results with outflow boundary conditions along the $x$-direction are robust to this choice, by comparing our default $n_0=4$ simulations against simulations with a higher $n_0=64$.  Further, we check if the many rounds (32) of binomial 1-2-1 current filtering that we employ in our simulations artificially affect our results. We do this by performing new simulations with a lower number (8) of rounds of current filtering.

Fig.~\ref{fig:spectra_ppc0_ntimes} shows the total energy spectra (left panel) and the bulk component (right panel). The overall conclusions of our paper ($\sim$100\,keV Maxwellian peak of electrons’ bulk spectrum, and a corresponding $\sim$100\,keV cut-off energy in X-ray spectrum) are determined primarily by the shape and location of the peak of the bulk spectrum. We note that these features are not different for either 8 rounds or 32 rounds of binomial current filtering.  The excess at low energies seen in the bulk spectrum at $\Gamma-1 \sim 2\times10^{-3}$ for the simulation with 32 rounds of filtering is due to the stochastic formation of a monster plasmoid. We also ran simulations with a much higher particle density of $n_0 = 64$, for both 8 and 32 rounds of filtering: our original results are found to be robust.

\section{Electron-ion energy exchange via Coulomb collisions}
\label{appendix:thermalization}
The coronal plasma is expected to be
approximately
collisionless  \citep{Goodman&Uzdensky_08}. However, the plasmas in PIC simulations differ from real astrophysical plasmas due to `coarse-graining', i.e., multiple particles are represented in the computer by a single macro-particle. This approach---while numerically beneficial---can artificially enhance the thermal coupling between electrons and ions via Coulomb collisions \citep{melzani_14}. 
Clearly, the artificial Coulomb coupling is not significant in our simulations, since the plasma ends up with two temperatures (hot ions and cold electrons; see Fig.~\ref{fig:energies_taucool_sigma}). This fact can also be seen using the following estimate for the Coulomb coupling timescale \citep{Spitzer_65, Stepney_83, Melzani+13},
\be \label{eq:t_relaxation}
t_{\rm ei} = \frac{3\mi\me c^3}{8\sqrt{2\pi}e^4}\frac{1}{{\mathcal{N}}}\left(\frac{k_{\rm B}T_{\rm e}}{\me c^2}+\frac{k_{\rm B}T_{\rm i}}{\mi c^2}\right)^{3/2}.
\ee
where $\mathcal{N}$ has units of particles per unit volume. The expression for $\mathcal{N}$ differs between two and three dimensions, due to the different volume element in the space of impact parameters: in 3D we have $\mathcal{N}=n \ln{\Lambda}$, where $n$ is the number of particles per unit volume, $\ln{\Lambda}=\ln{(b_{\rm max}/b_{\rm min})}$ is the Coulomb logarithm, and $b_{\rm min}$ and $b_{\rm max}$ are the minimum and maximum impact parameters for the Coulomb interaction, respectively. In 2D, instead, $\mathcal{N}=2\pi n_{\rm 2D}\left(\frac{1}{b_{\rm min}} - \frac{1}{b_{\rm max}}\right)\simeq2\pi n_{\rm 2D}/b_{\rm min}$, where $n_{\rm 2D}$ is the number of particles per unit area (as appropriate for a 2D simulation). The maximum impact parameter $b_{\rm max}$ is the electron Debye length, $\lambda_{\rm De}=\sqrt{k_{\rm B}T_{\rm e}/4\pi n_0 e^2}$. In real astrophysical plasmas, the minimum impact parameter $b_{\rm min}$ is the classical distance of closest approach. Instead, 
in PIC simulations it shall be taken as the characteristic size of macro-particles, which is comparable to the grid spacing $\delta$.

For our 2D simulations, we take a 2D density of $n_{\rm 2D}\simeq n_0=4$ particles per cell, temperatures $k_{\rm B}T_{\rm e}/\me c^2\ll k_{\rm B}T_{\rm i}/\mi c^2$, and the reduced mass ratio $\mi/\me=29$ of our simulations. Then we obtain $t_{\rm ei}\approx3\times10^5\,\omega_{\rm pe}^{-1}(k_{\rm B}T_{\rm i}/0.5\mi c^2)^{3/2}$. Note that this is much longer than the timescale for plasma advection out of the reconnection layer: $t_{\rm ei}/(L_{\rm x}/v_{\rm A}) \approx 60$ (for the most constraining case of our largest simulation with $L_{\rm x}/(\comp)=6720$, and $\sigmai=3$). This confirms that Coulomb energy exchange is weak
in our PIC simulations.

Marginally important Coulomb coupling is expected in real black-hole coronae with high magnetizations $\sigma>1$. It can be evaluated using the expression for $t_{\rm ei}$, but now with realistic plasma parameters. The ratio of the Coulomb and advection timescales may be written in the following form,
\begin{equation}
   \frac{t_{\rm ei}}{L_{\rm x}/v_{\rm A}}\sim \frac{\mi/\me}{10\,\tau_{\rm T} \ln \Lambda }\left(\frac{v_{\rm A}}{c}\right)^2\left(\frac{k_{\rm B}T_{\rm i}}{\mi c^2}\right)^{3/2}, \quad \frac{T_{\rm e}}{\me}\ll \frac{T_{\rm i}}{\mi}.
\end{equation}
Coulomb logarithm $\ln \Lambda\sim 20$ is determined by the characteristic plasma density $n=\tau_{\rm T}/h\sigma_{\rm T}$ where $h\approx 0.1(v_{\rm A}/c)L_{\rm x}$ is a fraction of the system size (a fraction of the black hole radius). Using $\tau_{\rm T}\sim 1$, $(v_{\rm A}/c)^2=\sigma/(1+\sigma)$, and $k_{\rm B}T_{\rm i}\sim (\sigma/4) \mi c^2$, we obtain a rough estimate $t_{\rm ei}/(L_{\rm x}/v_{\rm A})\sim 
\sigma^{5/2}/(1+\sigma)$. For a large $\sigma>1$, $t_{\rm ei}/(L_{\rm x}/v_{\rm A})$ exceeds unity, and the ions are not expected to pass their energy to electrons via Coulomb collisions. The Coulomb coupling becomes important at lower $\sigma\sim 1$. 

\section{Ion temperature anisotropy}\label{appendix:temp_anisotropy}

\begin{figure*}
\includegraphics[width=18cm]{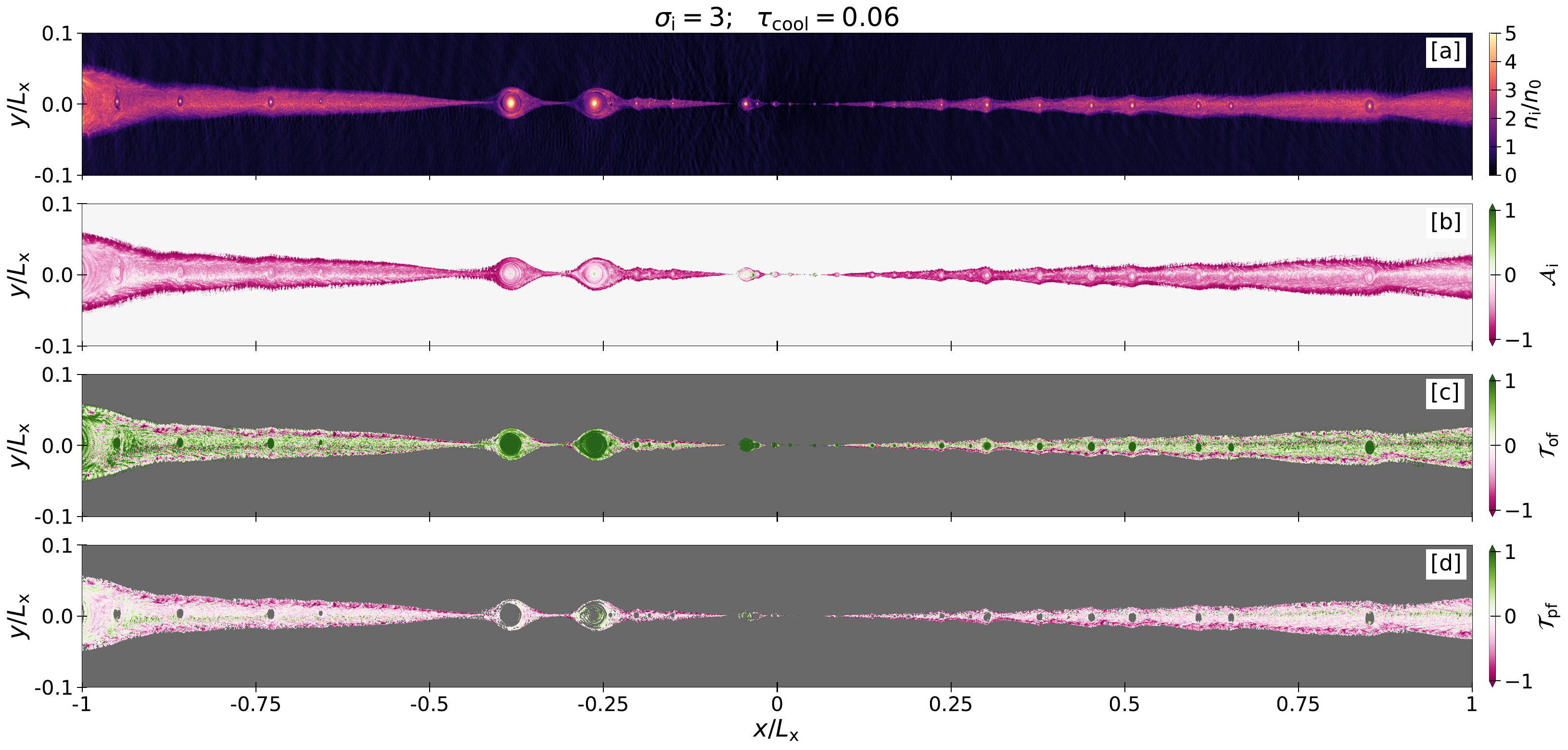}
\caption{Snapshot of the reconnection region at time $T\simeq 5\,\Lx/v_{\rm A}$ for the strongly-cooled ($\gcool=0.06$) run with $\sigmai=3$. We only show the region $|y|\le 0.1\,\Lx$ where reconnection occurs, where $\Lx/(\compe)=6720$. [a] Ion number density $n_{\rm i}$ in units of the initialized (upstream) number density $n_{\rm 0}$. [b] Ion temperature anisotropy ${\cal A}_{\rm i}=T_{\perp}^{\rm i}/T_{\parallel}^{\rm i}-1$, where $T_{\perp}^{\rm i}$ and $T_{\parallel}^{\rm i}$ are the temperatures of the ions perpendicular and parallel to the local magnetic field, respectively. [c] Threshold for triggering oblique firehose instability, ${\cal T}_{\rm of}=(\beta_{\parallel}+0.11){\cal A}_{\rm i}+1.4<0$, where $\beta_{\parallel}=8\pi n_{\rm i}k_{\rm B}T_{\parallel}^{\rm i}/B^2$ is the parallel ion plasma-$\beta$. [d] Threshold for triggering parallel firehose instability, ${\cal T}_{\rm pf}=(\beta_{\parallel}-0.59)^{0.53}{\cal A}_{\rm i}+0.47<0$.}
\label{fig:temp_anisotropy}
\end{figure*}

In this section, we examine the reconnection layer for the potential occurrence of collisionless ion velocity-space instabilities, that could transfer thermal energy between ions and electrons. We calculate in the reconnection region the ion temperature anisotropy parameter ${\cal A}_{\rm i}=T^{\rm i}_\perp/T^{\rm i}_{\parallel} - 1$, where $T^{\rm i}_\perp$ and $T^{\rm i}_{\parallel}$ are the temperatures of the ions perpendicular and parallel to the local magnetic field, respectively. The temperature elements are obtained as follows. The stress-energy tensor and the electromagnetic tensor are Lorentz boosted to the plasma comoving frame. A rotation matrix is then computed---using Rodrigues' rotation formula---so that the local comoving magnetic field gets oriented along $\hat{y}$. The Lorentz boosted stress-energy tensor is then rotated with this rotation matrix. The diagonal elements of thus obtained tensor correspond to the temperatures along different axes. In this case, $T_\parallel=T_{\rm yy}$ and $T_{\perp}=(T_{\rm xx}+T_{\rm zz})/2$.

In the presence of ion temperature anisotropy, the plasma can develop either ion-cyclotron/mirror instabilities for ${\cal A}_{\rm i}>0$, or firehose instability for ${\cal A}_{\rm i}<0$ \citep{Hall+79}. Fig.~\ref{fig:temp_anisotropy} shows a 2D map of ${\cal A}_{\rm i}$ along the reconnection layer (panel [b]). One can see that the ions' temperatures are fairly isotropic in reconnection plasmoids, that contain most of the particles (compare with the ion density in panel [a]). Furthermore, it is evident that the ion-cyclotron/mirror instabilities are non-operational throughout the layer (because ${\cal A}_{\rm i}\ngtr1$), precluding the possibility of ion-to-electron thermal energy transfer through this channel \citep[e.g.,][]{sironi_15}. On the other hand, the periphery of the reconnection layer exhibits regions (with ${\cal A}_{\rm i}<0$) possibly prone to the firehose instability. This is understandable, as the freshly injected particles in the periphery of the layer slide along the local magnetic field lines before eventually entering the plasmoids.

The threshold condition for the firehose instability  is given by \citep{Hellinger+06},
\be
\frac{T_{\perp}^{\rm i}}{T_{\parallel}^{\rm i}}-1 = \frac{a}{(\beta_{\parallel}-\beta_0)^b},
\ee
where $\beta_{\parallel}=8\pi n_{\rm i}k_{\rm B}T_{\parallel}^{\rm i}/B^2$ is the parallel ion plasma-$\beta$, and $a$, $b$ and $\beta_0$ are obtained from empirical fitting \citep{Gary+98, Samsonov+01, Hellinger+06, Bale+09}. The relevant modes in the case of ${\cal A}_{\rm i}<0$ are the parallel firehose \citep{Quest&Shapiro_96} and the oblique firehose \citep{Hellinger&Matsumoto_00}. Assuming isotropic Maxwellian electrons, the thresholds for these modes are
\be
{\cal T}_{\rm of}=(\beta_{\parallel}+0.11){\cal A}_{\rm i}+1.4<0 \ \ \ \ \ \ \ \ \ ({\rm oblique~firehose}),\\
{\cal T}_{\rm pf}=(\beta_{\parallel}-0.59)^{0.53}{\cal A}_{\rm i}+0.47<0 \ \ \ ({\rm parallel~firehose}).
\ee
We see from panel [c] of Fig.~\ref{fig:temp_anisotropy} that most of the reconnection region does not obey the condition ${\cal T}_{\rm of}<0$ required for the growth of the oblique firehose instability. On the other hand, ${\cal T}_{\rm pf}<0$ is satisfied along the periphery of the reconnection region---indicating that the parallel firehose instability may grow there (panel [d]). However, note that neither parallel nor oblique modes of firehose instabilities are seen to operate within reconnection plasmoids, that contain the majority of the particles. We can therefore confirm that ion velocity-space instabilities do not play any significant role in the transfer of thermal energy from ions to electrons, as our results indicate in the main text.

\section{Boundary conditions in radiative transfer calculation}\label{appendix:Monte_Carlo_bc}

\begin{figure} 
\includegraphics[width=8cm]{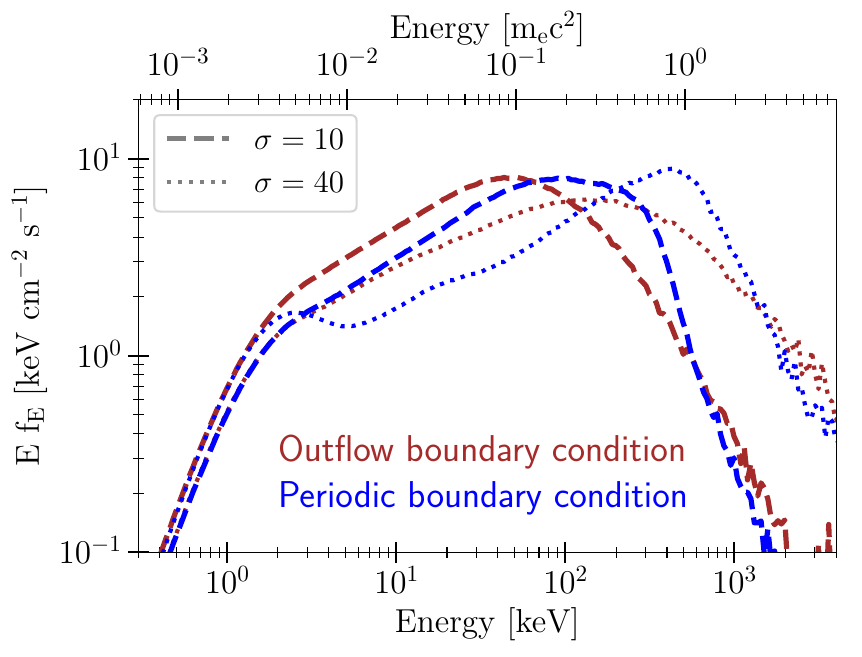}
\caption{Dependence on the choice of boundary conditions for the Monte Carlo radiative transfer calculation. We show electron-positron models with $\sigma=10$ (dashed) and $\sigma=40$ (dotted), with the same strength of radiative cooling losses ($\gcr=16$) and Compton amplification factor ($A=10$). The brown and blue curves are obtained by employing outflow and periodic boundary conditions, respectively.}
\label{fig:X-ray_spectra_bc}
\end{figure}

Our PIC simulations employ outflow boundary conditions in the $x$ direction of reconnection exhausts. In this Appendix, we demonstrate the effect of employing different boundary conditions for the Monte Carlo radiative transfer calculation. In our previous work (\citetalias{Sridhar+21c}), we extracted X-ray spectra from electron-positron simulations of reconnection in strongly magnetized plasma. There, the radiative transfer code assumed periodic boundaries in the $x$ direction. A more appropriate choice---as we have done in the main body of this paper---is to employ outflow boundaries for the Monte Carlo calculation, in analogy to the open boundaries of the PIC simulation. As shown in Fig.~\ref{fig:X-ray_spectra_bc}, at fixed magnetization, the X-ray spectrum peaks at higher energies in the case of periodic boundaries. This is expected, since in this case photons crossing the boundary can efficiently scatter off an oppositely-directed flow, thus increasing their energy. 

%%%%%%%%%%%%%%%%%%%%%%%%%%%%%%%%%%%%%%%%%%%%%%%%%%

% Don't change these lines
\bsp	% typesetting comment
\label{lastpage}
\end{document}